\documentclass[10pt,letterpaper]{article}
\usepackage[margin=3cm]{geometry}

\usepackage{amsmath,amssymb}

\usepackage{changepage}

\usepackage[utf8x]{inputenc}

\usepackage{textcomp,marvosym}

\usepackage{cite}

\usepackage{nameref,hyperref}

\usepackage{microtype}
\DisableLigatures[f]{encoding = *, family = * }

\usepackage[table]{xcolor}

\usepackage{array}

\newcolumntype{+}{!{\vrule width 2pt}}

\newlength\savedwidth

\usepackage[aboveskip=1pt,labelfont=bf,labelsep=period,justification=raggedright,singlelinecheck=off]{caption}

\makeatletter
\renewcommand{\@biblabel}[1]{\quad#1.}
\makeatother

\usepackage{lastpage,fancyhdr,graphicx}
\usepackage{epstopdf}
\fancyheadoffset[L]{2.25in}
\fancyfootoffset[L]{2.25in}
\usepackage{epstopdf}

\usepackage{caption}
\usepackage{subcaption}

\usepackage{svg}

\begin{document}
\vspace*{0.2in}

\begin{flushleft}
{\Large
\textbf\newline{Correlations from structure and phylogeny combine constructively in the inference of protein partners from sequences} % Please use "sentence case" for title and headings (capitalize only the first word in a title (or heading), the first word in a subtitle (or subheading), and any proper nouns).
}
\newline
\\
Andonis Gerardos\textsuperscript{1,2,\textcurrency},
Nicola Dietler\textsuperscript{1,2},
Anne-Florence Bitbol\textsuperscript{1,2*}
\\
\bigskip
\textbf{1} Institute of Bioengineering, School of Life Sciences, École Polytechnique Fédérale de Lausanne (EPFL), CH-1015 Lausanne, Switzerland\\
\textbf{2} SIB Swiss Institute of Bioinformatics, CH-1015 Lausanne, Switzerland
\bigskip

% Insert additional author notes using the symbols described below. Insert symbol callouts after author names as necessary.
% 
% Remove or comment out the author notes below if they aren't used.
%
% Primary Equal Contribution Note
%\Yinyang These authors contributed equally to this work.

\textcurrency Current address: Aix Marseille Univ, CNRS, CINAM, Marseille, France % change symbol to "\textcurrency a" if more than one current address note
% \textcurrency b Insert second current address 
% \textcurrency c Insert third current address

% Use the asterisk to denote corresponding authorship and provide email address in note below.
* anne-florence.bitbol@epfl.ch

\end{flushleft}
\section*{Abstract}
Inferring protein-protein interactions from sequences is an important task in computational biology. Recent methods based on Direct Coupling Analysis (DCA) or Mutual Information (MI) allow to find interaction partners among paralogs of two protein families. Does successful inference mainly rely on correlations from structural contacts or from phylogeny, or both? Do these two types of signal combine constructively or hinder each other? To address these questions, we generate and analyze synthetic data produced using a minimal model that allows us to control the amounts of structural constraints and phylogeny. We show that correlations from these two sources combine constructively to increase the performance of partner inference by DCA or MI. Furthermore, signal from phylogeny can rescue partner inference when signal from contacts becomes less informative, including in the realistic case where inter-protein contacts are restricted to a small subset of sites. We also demonstrate that DCA-inferred couplings between non-contact pairs of sites improve partner inference in the presence of strong phylogeny, while deteriorating it otherwise. Moreover, restricting to non-contact pairs of sites preserves inference performance in the presence of strong phylogeny. In a natural data set, as well as in realistic synthetic data based on it, we find that non-contact pairs of sites contribute positively to partner inference performance, and that restricting to them preserves performance, evidencing an important role of phylogeny.

% Please keep the Author Summary between 150 and 200 words
% Use first person. PLOS ONE authors please skip this step. 
% Author Summary not valid for PLOS ONE submissions.   
\section*{Author summary}

In protein sequence data, the amino acid usages at different sites of a protein or of two interacting proteins can be correlated because of functional constraints. For instance, the need to maintain physico-chemical complementarity among two sites that are in contact in the three-dimensional structure of a protein complex causes such correlations. However, correlations can also arise due to shared evolutionary history, even in the absence of any functional constraint. While these phylogenetic correlations are known to obscure the inference of structural contacts, we show, using controlled synthetic data, that correlations from structure and phylogeny combine constructively to allow the inference of protein partners among paralogs using just sequences. We also show that pairs of amino acids that are not in contact in the structure have a major impact on partner inference in a natural data set and in realistic synthetic ones. These findings explain the success of methods based on pairwise maximum-entropy models or on information theory at predicting protein partners from sequences among paralogs.

%\linenumbers

% Use "Eq" instead of "Equation" for equation citations.
\section*{Introduction}

Most cellular processes are carried out by interacting proteins. Thus, mapping protein-protein interactions is a crucial goal. Since high-throughput experiments remain challenging~\cite{Rajagopala14}, it is interesting to exploit the growing amount of available sequence data to identify candidate protein-protein interaction partners. The amino-acid sequences of interacting proteins are correlated, both because of evolutionary constraints arising from the need to maintain physico-chemical complementarity between amino acids that are in contact in the three-dimensional structure of protein complexes, and because of shared evolutionary history. On the one hand, correlations from structural contacts have received substantial interest, both within single proteins and across interacting protein partners. Global statistical models~\cite{Lapedes99,Burger08} using the maximum entropy principle~\cite{Jaynes57} and designed to match the one- and two-body statistics of natural sequence data, often called Direct Coupling Analysis (DCA)~\cite{Weigt09}, have allowed to determine three-dimensional protein structures from sequences~\cite{Marks11,Morcos11,Sulkowska12}, to analyze mutational effects~\cite{Dwyer13,Cheng14,Cheng16,Figliuzzi16}, protein evolution~\cite{delaPaz20} and conformational changes~\cite{Morcos13,Malinverni15}, to design proteins~\cite{Russ20}, to find residue contacts between known interaction partners~\cite{Weigt09,Procaccini11,Baldassi14,Ovchinnikov14,Hopf14,Tamir14,dosSantos15,Feinauer16}, and to predict interaction partners among paralogs~\cite{Bitbol16,Gueudre16} and protein-protein interaction networks~\cite{Cong19,Green21} from sequence data. 
On the other hand, correlations arise in protein sequences due to their common evolutionary history, i.e. phylogeny~\cite{Casari95,Halabi09,Qin18}, even in the absence of structural constraints. Functionally related~\cite{Fryxell96} and interacting~\cite{Goh00} protein families tend to have similar phylogenies. This can arise from global shared evolutionary pressures on interacting partners, resulting in similar evolutionary rates~\cite{Hakes07,Juan08,Kann09,Lovell10,Swapna12}, and from mere shared evolutionary history, including common timing of speciations and gene duplications~\cite{Lovell10}. Accordingly, methods based on sequence similarity, e.g. Mirrortree~\cite{Pazos01,Jothi05,Bradde10,Ochoa10,Ochoa15}, or on the simultaneous presence and absence of genes, e.g. phylogenetic profiling~\cite{Pellegrini99,Croce19,Moi20} allow to predict which protein families interact. Mutual information (MI), which includes all types of statistical dependence between the sequences of interacting partners, slightly outperforms DCA at predicting interaction partners among paralogs~\cite{Bitbol18}. While DCA allows to infer interaction partners among paralogs in synthetic data that only comprises correlations from contacts~\cite{Gandarilla20}, good performance is also obtained at this task by DCA and MI in synthetic data that only includes phylogenetic correlations~\cite{Marmier19}. Therefore, correlations from contacts and from phylogeny are both useful to predict protein-protein interaction partners among paralogs. This stands in contrast with the identification of structural contacts by DCA~\cite{Weigt09,Marks11,Qin18,Vorberg18,RodriguezHorta19,RodriguezHorta21}, where phylogenetic correlations obscure structural ones, motivating the use of phylogeny corrections~\cite{Lichtarge96,Dunn08}, such as the Average Product Correction~\cite{Ekeberg13,Hockenberry19}, reweighting close sequences~\cite{Marks11,Morcos11,Malinverni20,Hockenberry19}, and Nested Coevolution~\cite{Colavin22}. 

Recently, deep learning has brought major advances to the computational prediction of protein structures~\cite{Jumper21}, and these approaches have been extended to the prediction of protein-protein interactions and of protein complex structures~\cite{Humphreys21,EvansPreprint,Bryant22}, outperforming DCA-based ones~\cite{Humphreys21}. Coevolution between candidate partners is an important ingredient of these new methods. For each candidate pair of proteins, they start by constructing a paired or partially paired multiple sequence alignment of homologs, using species information, orthology, and genome proximity in prokaryotes~\cite{Cong19,Green21,Humphreys21,EvansPreprint,Bryant22}. These paired alignments are used as input of the deep neural networks, and their quality impacts that of the final predictions~\cite{Humphreys21,Bryant22}. Therefore, DCA- and MI-based methods to infer partners among paralogs from sequences, such as those introduced in~\cite{Bitbol16,Gueudre16,Bitbol18} could be employed to enrich and improve these input alignments. 

How do DCA- and MI-based methods to infer protein partners among paralogs of two protein families perform in the presence of both phylogeny and structural contacts? Does successful inference mainly rely on one or the other of these two sources of correlations? Do they combine constructively or hinder each other? What changes when one dominates over the other? Answering these questions is important to understand the performance of DCA- and MI-based methods to infer protein partners in natural data, and should help to develop new methods that combine information from both phylogeny and contacts in an optimal way. However, a substantial challenge is that real data sets contain a given mix of these signals, which is unknown and difficult to disentangle. Therefore, to address these questions, we generate synthetic data in a minimal model that allows us to control the amounts of structural constraints and phylogeny. We also consider a data set of natural sequences, as well as synthetic data generated using models inferred on this natural data. Our focus is on predicting partners among paralogs, starting from a training set of known interaction partners, but our methods can be extended to the case where there is no training set via an Iterative Pairing Algorithm (IPA)~\cite{Bitbol16, Bitbol18}.

We find that correlations from structural contacts and from phylogeny add constructively in partner inference by DCA or MI. Furthermore, the signal from phylogeny can rescue partner inference in regimes of relatively weak selection and in the realistic case where inter-protein contacts are restricted to a small subset of sites. We show that DCA-inferred couplings between non-contact sites improve partner inference in the presence of strong phylogeny, while deteriorating it otherwise, and that they suffice to obtain good performance in the presence of strong phylogeny. In a natural data set, as well as in realistic synthetic data, we find that non-contact pairs of sites contribute positively to the performance of partner inference among paralogs, and that restricting to them preserves performance, evidencing an important role of phylogeny.

\section*{Model and methods}  

\subsection*{Modeling structural constraints with Potts models}

\paragraph{General approach.} We model the constraints stemming from the physicochemical complementarity of amino acids that are in contact in the three-dimensional structure of protein complexes by pairwise interactions in a Potts model. We consider concatenated sequences composed of two interacting partners A and B with respective lengths $L_A$ and $L_B$. We denote by $\alpha_i\in\{1,\dots,q\}$ the state of site $i\in\{1,\dots,L_A+L_B\}$, where $q$ is the number of possible states. The Hamiltonian of a concatenated sequence $\vec{\alpha}=(\alpha_1,...,\alpha_{L_A+L_B})$ reads: 
\begin{equation}
H(\vec{\alpha})=-\sum_{i=1}^{L_A+L_B}h_{i}(\alpha_i)-\sum_{j=1}^{L_A+L_B}\sum_{i=1}^{j-1}e_{ij}(\alpha_i,\alpha_j)\,,
\label{maxent}
\end{equation}
where fields $h_i$ yield conservation, while (direct) couplings $e_{ij}$ model pairwise interactions. 
Pairwise maximum entropy inference (DCA) yields the Potts Hamilitonian in Eq.~\ref{maxent}~\cite{Weigt09,Cocco18}. 

\paragraph{Minimal model.} 
In our minimal model, sequences are strings of binary variables represented by ``Ising spins'' taking values $-1$ or 1 ($q=2$). Equivalently, one could take values 0 or 1, which is more usual for proteins~\cite{Halabi09,Dahirel11,Mann14}, but we choose the spin convention to make the link with statistical physics~\cite{Gandarilla20}.
These spins are coupled via uniform ferromagnetic couplings, set to unity, on all edges of an Erd\H{o}s-Rényi random graph, all other couplings being zero, and all fields being zero. In the Ising (or zero-sum) gauge~\cite{Ekeberg14}, employing Ising notations, i.e. $h(\alpha_i)=h_i\alpha_i$ and $e_{ij}(\alpha_i,\alpha_j)=J_{ij}\alpha_i\alpha_j$, and then implementing our choices $h_i=0$ for all $i$ and $J_{ij}=1$ if $(i,j)\in {\cal E}$, where ${\cal E}$ is the set of edges of the Erd\H{o}s-Rényi graph, and $J_{ij}=0$ otherwise, the Hamiltonian in Eq.~\ref{maxent} can be simplified to
\begin{equation}
H(\vec \alpha) =-\sum_{i=1}^{L_A+L_B}h_{i}\alpha_i - \sum_{j=1}^{L_A+L_B}\sum_{i=1}^{j-1}J_{ij}\alpha_i\alpha_j= - \sum_{(i,j)\in {\cal E}} \alpha_i \alpha_j\; .
\label{miniHam}
\end{equation}
Here, the sequence $\vec \alpha = (\alpha_1,...,\alpha_{L_A+L_B})\in \{\pm 1\}^{L_A+L_B}$ is a string of $L_A+L_B$ Ising spins. Importantly, the Erd\H{o}s-Rényi graph is fixed throughout, as it models the set of contacts of two given interacting protein families assembling into a specific complex structure. For simplicity we take $L_A=L_B=L$. We consider an Erd\H{o}s-Rényi graph with $2L=200$ vertices, where any two vertices are connected with probability $p=0.02$. 
Because inter-protein contacts are generally sparser than intra-protein ones, we also study other graphs satisfying this constraint. 

\paragraph{Models inferred from real data.} We also generate more realistic synthetic data from Potts models inferred from two different natural sequence data sets. The first one is composed of $23,633$ pairs of natural sequences of interacting histidine kinases (HK) and response regulators (RR) from the P2CS database~\cite{Barakat09,Barakat11}. The second one comprises 17,950 pairs of ABC transporter proteins homologous to the \textit{Escherichia coli} MALG-MALK pair of maltose and maltodextrin transporters~\cite{Ovchinnikov14,Bitbol16}. In both of these natural data sets, interacting partners are determined using proximity in the genome (either using annotations of the P2CS database or following the approach from Ref.~\cite{Ovchinnikov14}), allowing us to assess partner inference performance in these natural data sets as well as in synthetic ones. In protein sequences, there are $q=21$ states, namely the 20 natural amino acids and the alignment gap. We use state-of-the-art methods that have good generative properties, namely bmDCA~\cite{Figliuzzi18,Russ20} and arDCA~\cite{Trinquier21}. In practice, we employ bmDCA with its default parameters for $q=21$, and with default parameters except $t_{wait,0}=1000$ and $\Delta t_0=100$ for $q=2$ (motivated by the faster equilibration observed for $q=2$). For arDCA, we use default parameters, apart from the reweighting parameter $\theta=0.2$ (chosen to match the bmDCA value) and the regularization strengths $\lambda_J = 2\times10^{-4}$ and $\lambda_h = 10^{-5}$ for $q=21$ or $\lambda_J = 2\times10^{-3}$ and $\lambda_h = 10^{-4}$ for $q=2$ (chosen to reproduce one- and two-body frequencies well, see Figs.~\ref{fig:fij1},~\ref{fig:fij2} and~\ref{fig:two-body-freq}).

\subsection*{Generating synthetic data with controlled amounts of structural constraints and phylogeny}

\paragraph{General approach.} 
We generate synthetic data using Markov Chain Monte Carlo sampling along a phylogenetic tree~\cite{Qin18,Vorberg18}, employing the Potts model Hamiltonian $H$ in Eq.~\ref{maxent} (for $q=21$) or Eq.~\ref{miniHam} (for $q=2$) to model structural constraints. As we focus on pairs of protein families with given structures, we assume that the ancestral protein complex already had the same structural constraints, and we take as our ancestral concatenated sequence AB an equilibrium sequence under the Hamiltonian $H$ at sampling temperature $T$. We then simulate evolution along the chosen phylogenetic tree (see below): random mutations are proposed at sites chosen uniformly at random, independently on each branch of the tree. Each proposed mutation is accepted with a probability $p$ given by the Metropolis criterion at sampling temperature $T$:
\begin{equation}
p=\min\left[1,\exp\left(-\frac{\Delta H}{T}\right)\right]\,,
\label{accept}
\end{equation}
where $\Delta H$ is the difference between the value of $H$ after the mutation and before it.
This models natural selection for maintaining structure~\cite{Gandarilla20}. Indeed, all mutations that decrease $H$ are accepted ($p=1$), while those that increase $H$ can be rejected ($p<1$), and will generally be rejected if $T$ is small.  

\begin{figure}[h]
	\centering
	\includegraphics[width=\textwidth]{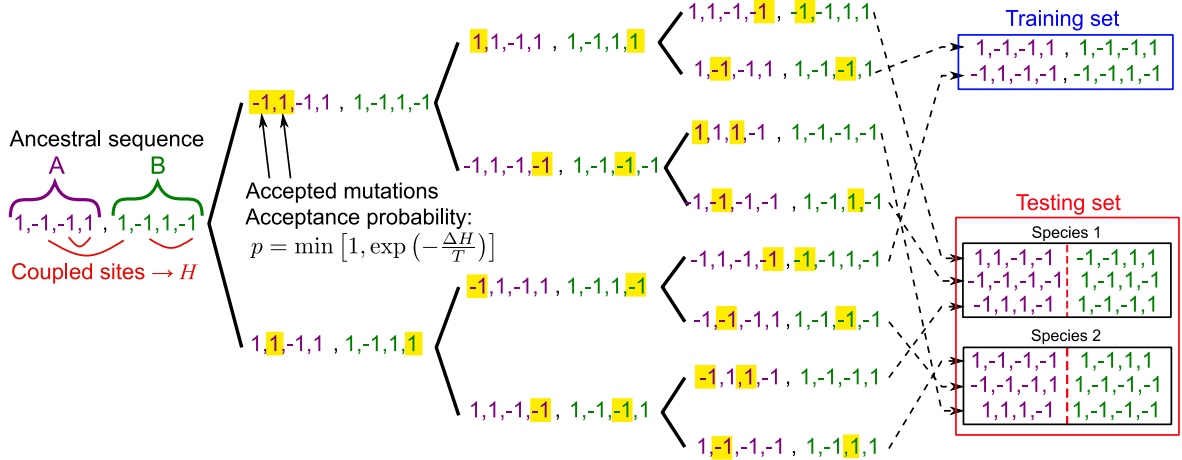}
	\vspace{0.2cm}
	\caption{\textbf{Construction of a synthetic data set in the minimal model.} A binary branching tree provides a minimal phylogeny. Structural constraints are represented by pairwise couplings on the edges of an Erd\H{o}s-Rényi graph (Hamiltonian $H$ in Eq.~\ref{miniHam}). Starting from an ancestral concatenated sequence AB of $2L$ Ising spins (here $L=4$), a series of $n$ duplication events (``generations'', here $n = 3$) are performed. Between these duplications, and independently on each branch of the tree, $\mu$ mutations take place (here $\mu=2$, and mutated sites are highlighted). Each mutation is an accepted spin flip at one site, and the acceptance criterion is given by Eq.~\ref{accept}. This process yields $2^n = 8$ chains AB at the tree leaves. This data is then randomly split into a training set and a testing set. In the testing set, groups of $m$ sequences modeling species are randomly formed (here $m = 3$). Next, the evolutionary and functional pairings between chains A and B are blinded within each species of the testing set.}
	\label{fig:phylo_tree}
\end{figure}

\paragraph{Minimal model.} In our minimal model, a simple phylogeny is introduced via a binary branching tree with a fixed number $n$ of ``generations'' (duplication events) and a fixed number $\mu$ of accepted mutations on each branch (between two subsequent duplication events; see Fig.~\ref{fig:phylo_tree}). It gives rise to $2^n$ concatenated sequences AB on the leaves of the tree, which constitute a synthetic data set of paired sequences, where partners A and B evolved together along the tree. In practice, we choose $n=10$, and thus $2^n=1024$, ensuring that inference works well without phylogeny~\cite{Gandarilla20}, and is computationally fast.

\paragraph{Controlling the importance of structural constraints and phylogeny.} Tuning the sampling temperature $T$ and the number $\mu$ of mutations per branch allows us to control the relative importance of structural constraints and phylogeny. First, Eq.~\ref{accept} shows that when $T\to 0$, mutations that increase the value of $H$ are all rejected, yielding strict selection for structure. Conversely, when $T\to\infty$, all mutations are accepted, so that all correlations in the data arise from phylogeny (and finite-size noise). Second, if $\mu$ is small, then all sequences resemble the ancestral one, yielding extreme phylogenetic correlations. If $\mu$ is very large, even sister sequences arising from the last branching event become independent~\cite{Marmier19}, leaving only correlations from structure.

\paragraph{Pure structural constraints limit and pure phylogeny limit.} First, to consider the limiting case that only involves structural constraints, independent equilibrium sequences are generated using the Markov Chain Monte Carlo sampling scheme explained above (see Eq.~\ref{accept}). Each sequence is generated starting from a different initial random sequence~\cite{Gandarilla20}. The equilibration time is determined by the convergence of the Hamiltonian value, see Fig.~\ref{fig:magnetisation_against_mutations} (note that the convergence of the absolute magnetization correlation function gives similar results~\cite{Gandarilla20}). Second, to consider the limiting case that only involves phylogeny, synthetic sequences are evolved along a phylogenetic tree, and all proposed mutations are accepted, simulating neutral evolution, where mutations have no fitness effect. Note that with structural constraints, when $T\to\infty$, all proposed mutations are also accepted since $|\Delta H|\ll T$, despite the fact that they may be deleterious ($\Delta H>0$). 

\paragraph{Models inferred from real data.} We employ generative models inferred on natural paired sequences (see above) to generate realistic synthetic data either without or with phylogeny. While bmDCA infers a Potts model, arDCA directly infers the distribution of probability of sequences~\cite{Trinquier21}. First, to generate contact-only data, we employ a Markov Chain Monte Carlo procedure for bmDCA (equilibrium is considered reached after $10^6$ accepted mutations for $q=21$, or $10^5$ for $q=2$), while we directly sample independent sequences from the inferred distribution for arDCA. Second, to generate data that incorporates both phylogeny and contacts~\cite{Vorberg18}, we employ a tree inferred on the data set of natural paired sequences via FastTree2 \cite{Price10}. As the length $b$ of a branch gives the mutation probability per site along it~\cite{Price10}, we generate data by making $\lfloor b\times(L_A+L_B)\rfloor$ mutations on a branch of length $b$.

\subsection*{Inference problem}

\paragraph{Question.} Given two protein families A and B that interact, and starting from a training set of known AB partners, we aim to find, in each species of a testing set (see Fig.~\ref{fig:phylo_tree}), which specific proteins A and B are evolutionary and functional partners. We assume for simplicity that there is a strict one-to-one pairing between each A and its partner B. 

\paragraph{Species in the minimal model.} In our minimal model, we randomly group concatenated AB sequences into sets of equal size $m$, representing species. The $m$ different sequences A (or B) within a species represent paralogs. This minimal model, where species contain random assortments of sequences, is realistic if exchange between species (horizontal gene transfer) is sufficiently frequent. As in Ref.~\cite{Marmier19}, we compare this random-species model to more realistic ones (see Fig.~\ref{fig:arbre_alternatif}), and qualitative conclusions are not affected, although the minimal model yields higher phylogenetic signal. In the testing set, within each species, we blind the pairings of the chains A and B. We then aim to infer these pairings, i.e. to recover for each A chain its evolutionary and functional partner, which is the B chain that coevolved with it.

\paragraph{Species in the model inferred from natural data.} In our more realistic model where synthetic sequences are generated employing Hamiltonians and phylogenies inferred from natural sequence data, we rely on the inferred tree to define species. On each leaf of the inferred tree lies a natural paired sequence, coming from a given species. When generating data along this tree, we put the generated sequence on a given leaf of the tree into the species associated to this leaf in the natural data. The distribution of the number of sequence pairs per species in the synthetic data then exactly matches that of the natural data, and sequences in each species have the same positions in the tree as in the natural data. The inference question is then asked in the exact same way as in the minimal model.

\subsection*{Inference methods}

\paragraph{Training set statistics.} The statistics of the training set of paired chains AB, of total length $L_A+L_B$, are described using the empirical one-site frequencies of each state $\alpha_i$ at each site $i\in\{1,\dots,L_A+L_B\}$, denoted by $f_i(\alpha_i)$, and the two-site frequencies of occurrence of each ordered pair of states $(\alpha_i,\alpha_j)$ at each ordered pair of sites $(i,j)$, denoted by $f_{ij}(\alpha_i,\alpha_j)$. Covariances are computed as $C_{ij}(\alpha_i,\alpha_j)=f_{ij}(\alpha_i,\alpha_j)-f_i(\alpha_i)f_j(\alpha_j)$. When we employ mean-field DCA (mfDCA) and mutual information (MI), pseudocounts with weights denoted by $\lambda$ and defined as in Refs.~\cite{Bitbol16,Bitbol18,Marmier19} are introduced~\cite{Procaccini11,Marks11,Morcos11,Bitbol18}.  The value $\lambda=0.5$ is usually employed in mfDCA~\cite{Marks11,Morcos11,Bitbol16}, while smaller values have proved better for MI~\cite{Bitbol18,Marmier19}. Thus, we always take $\lambda=0.5$ for mfDCA and $\lambda=0.01$ for MI. However, we do not employ any phylogenetic reweighting (except where noted) because our aim is to investigate the effect of phylogeny, and also because this reweighting has very little impact on the inference of partners~\cite{Bitbol16}. 

\paragraph{DCA-based inference method.} 
DCA is based on building a global statistical model consistent with the empirical one- and two-body frequencies of the training set~\cite{Weigt09,Morcos11,Marks11,Cocco18}, through the maximum entropy principle~\cite{Jaynes57}. This results in a probability of observing a given sequence reading~\cite{Cocco18}:
\begin{equation}
P(\alpha_1,\dots,\alpha_{L_A+L_B})=\frac{\exp\left[-H\right]}{Z}\,,
\label{Boltz}
\end{equation} 
where the Hamiltonian $H$ is given by Eq.~\ref{maxent}, i.e. by the Potts model, and $Z$ is a normalization constant.
Inferring the couplings and the fields that appropriately reproduce the empirical covariances is a difficult problem~\cite{Nguyen17}. Within the mean-field approximation (mfDCA), which we employ for partner inference as in~\cite{Bitbol16,Marmier19,Gandarilla20}, inferred coupling strengths can be simply approximated by $\hat{e}_{ij}(\alpha_i,\alpha_j)=-C^{-1}_{ij}(\alpha_i,\alpha_j)$ in the reference-state gauge~\cite{Plefka82,Morcos11,Marks11}. One then makes a gauge change to the zero-sum (or Ising) gauge~\cite{Ekeberg13,Bitbol16,Marmier19}, which attributes the smallest possible fraction of the energy to the couplings, and the largest possible fraction to the fields~\cite{Weigt09,Ekeberg13}. 

The effective interaction energy $E_{AB}$ of each possible pair AB in the testing set is given by
%, constructed by concatenating a chain A and a chain B from the same species,
\begin{equation}
E_{AB}=-\sum_{i=1}^{L_A}\sum_{j=L_A+1}^{L_A+L_B} \hat{e}_{ij}(\alpha_i^A,\alpha_j^B)\,.
\label{energy}
\end{equation}
In real proteins, approximately minimizing such a score has proved successful at predicting interacting partners~\cite{Bitbol16}. Note that we only sum over inter-protein pairs of sites (i.e. involving one site in A and one in B), and that we do not include fields, because we focus on interactions between A and B.

\paragraph{MI-based inference method~\cite{Bitbol18}.} The pointwise mutual information (PMI) of a pair of states $(\alpha_i,\alpha_j)$ at a pair of sites $(i,j)$ is defined from the empirical one and two-body frequencies of the training set as~\cite{Fano61,Church90,Role11}:
\begin{equation}
\textrm{PMI}_{ij}(\alpha_i,\alpha_j)=\log\left[\frac{ f_{ij}(\alpha_i,\alpha_j)}{ f_i(\alpha_i) f_j(\alpha_j)}\right]\,.
\label{PMI}
\end{equation}

A pairing score $S_\mathrm{AB}$ for each possible pair AB in the testing set can then be defined as the sum of the PMIs of the inter-protein pairs of sites of this concatenated chain AB (i.e. those involving one site in chain A and one site in chain B):
\begin{equation}
S_\mathrm{AB}=\sum_{i=1}^{L_A}\sum_{j=L_A+1}^{L_A+L_B}\textrm{PMI}_{ij}(\alpha_i^A,\alpha_j^B)\,.
\label{SAB}
\end{equation}

\paragraph{From scores to partner prediction.} Our goal is to find the best 1-to-1 mapping of putative partners A-B in each species of the testing set. We assign a score to each possible partner of interaction using Eq.~\ref{energy} or Eq.~\ref{SAB}, and select the one-to-one assignment which optimizes the sum of scores for all chosen pairs by solving the corresponding linear assignment problem~\cite{Kuhn55,Munkres57,HungAlg,Virtanen20}. 

\subsection*{Code availability}

The code associated to this paper is available at: \url{https://doi.org/10.5281/zenodo.6480583}.

\section*{Results}

\subsection*{Correlations from structural contacts and from phylogeny both contribute to the performance of partner inference}

In order to understand the origin of the performance of partner inference from protein sequences observed on real data using DCA~\cite{Bitbol16} and MI~\cite{Bitbol18}, we construct synthetic data sets from a minimal model where the contributions of structural contacts and phylogeny can be tuned via the number $\mu$ of mutations per branch of the tree and the sampling temperature $T$ (see Methods). How do these two parameters impact the performance of partner inference? Addressing this question will provide insight into the interplay of correlations from structural contacts and from phylogeny in partner inference. 

\begin{figure}[h]
	\centering
	\includegraphics[width=0.7\textwidth]{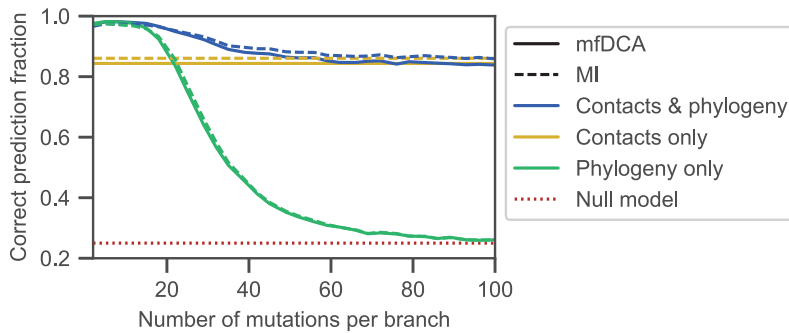}
	\caption{\textbf{Impact of mutation number on partner inference performance.} The fraction of correctly predicted partner pairs is shown versus the number $\mu$ of mutations per branch of the tree for the minimal model incorporating constraints both from contacts and from phylogeny. Specifically, an ancestral chain AB of $2L=200$ spins is evolved along a binary branching tree (see Fig. \ref{fig:phylo_tree}) with $n=10$ generations, and $\mu$ mutations per branch, yielding $2^{10}=1024$ pairs AB, at sampling temperature $T=5$ under the Hamiltonian in Eq.~\ref{miniHam} on an Erd\H{o}s-Rényi graph with $p=0.02$. The limiting cases with only contacts and only phylogeny are also shown for comparison. The first one corresponds to independent equilibrium sequences at $T=5$ under the Hamiltonian in Eq.~\ref{miniHam} on the same graph. The second one corresponds to neutral evolution on the same binary branching tree. Each of these three data sets is randomly split into a training set of 400 chains AB and a testing set of 624 chains AB. The latter are randomly divided in 156 species of 4 chains AB each, and pairings between A and B chains are blinded in each species. Partnerships are then predicted using the scores in Eq.~\ref{energy} (mfDCA) or Eq.~\ref{SAB} (MI). Each point is averaged over 20 generated data sets, and 20 random choices of the training set for each of them. The null model shows the expectation of the correct prediction fraction if pairs are made randomly within each species. }
	\label{fig:partners_mutation}
\end{figure}

\paragraph{Impact of the number of mutations per branch.} Fig~\ref{fig:partners_mutation} shows the impact of varying the number $\mu$ of mutations per branch of the tree on the performance of partner inference, measured via the fraction of correctly predicted partner pairs (recall that each protein A in the testing set is paired with one partner B within its species, see Methods). 
In Fig~\ref{fig:partners_mutation}, when $\mu\lesssim15$, the performance of partner inference in our data set that incorporates both structural contacts and phylogeny approaches that of a data set that only involves phylogeny (see Methods and Ref.~\cite{Marmier19}). Indeed, similarities between related sequences are large for small $\mu$, yielding dominant phylogenetic correlations.
Conversely, when $\mu\gtrsim 70$, Fig~\ref{fig:partners_mutation} shows that the performance of partner inference in our data set including both ingredients approaches the one obtained with only structural contacts (see Methods and Ref.~\cite{Gandarilla20}). Indeed, when $\mu$ becomes large enough, similarities due to phylogeny vanish even between closest relatives, and all sequences become effectively independent. More precisely, the number of differences between two sister sequences AB arising from the last duplication events is about $2\mu$ (exactly $2\mu$ if all mutations affect different sites), and if it is of the same order as the total sequence length $2L$, or larger, i.e. if $\mu\gtrsim L=100$ here, then even sister sequences lose all phylogenetic correlations.
Accordingly, Fig~\ref{fig:partners_mutation} shows that for $\mu\gtrsim 90$, performance in the pure-phylogeny data set drops to the chance expectation (``null model''), which corresponds to making random one-to-one pairings of sequences A and B within each species.  
Fig~\ref{fig:partners_mutation} also demonstrates that DCA and MI yield similar performance for partner inference, with MI becoming slightly better when phylogeny is not too strong, consistently with Refs.~\cite{Bitbol18,Marmier19}. 

Importantly, Fig~\ref{fig:partners_mutation} shows that partner inference performance in our data set including contacts and phylogeny is better than for both limiting data sets. Therefore, partner inference is made more robust by the fact that correlations from contacts and from phylogeny both contribute. Depending on how strong phylogeny is (i.e., here, on how small $\mu$ is), the dominant ingredient is either contacts or phylogeny, but in the generic case, these two signals add constructively to increase performance.

\begin{figure}[h]
	\centering
	\includegraphics[width=\textwidth]{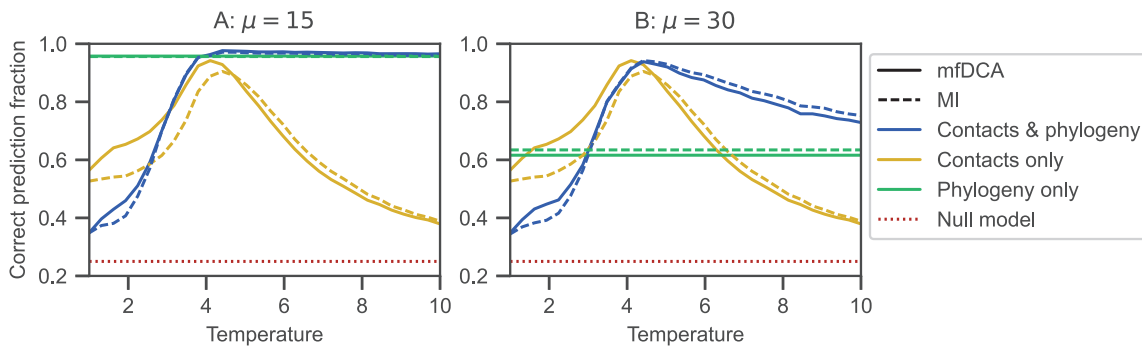}
	\caption{\textbf{Impact of sampling temperature on partner inference performance.} The fraction of correctly predicted partner pairs is shown versus the sampling temperature $T$ for the minimal model incorporating constraints both from contacts and from phylogeny. Data generation and inference are performed exactly as in Fig.~\ref{fig:partners_mutation}, using the same parameters and the same graph for contacts. However, here, $T$ is varied and $\mu=15$ (panel A) or $\mu=30$ (panel B). The limiting cases with only contacts and only phylogeny are also shown for comparison. The null model shows the expectation of the correct prediction fraction if pairs are made randomly within each species. }
	\label{fig:partners_temp}
\end{figure}

\paragraph{Impact of sampling temperature.} Fig~\ref{fig:partners_temp} shows the impact of varying the sampling temperature $T$ on the performance of partner inference for two different values of $\mu$, one where phylogeny dominates, $\mu=15$, and one where contacts and phylogeny both have an important contribution, $\mu=30$ (see Fig~\ref{fig:partners_mutation}). The sampling temperature $T$ impacts inference because $1/T$ is a proxy for the strength of selection on structural contacts (see Methods). It is analogous to the effective temperature at which foldable sequence have been selected in sequence space by evolution~\cite{Morcos14}. In addition, in our minimal model, a phase transition between a ferromagnetic (ordered) phase where all spins tend to align and a paramagnetic (disordered) phase occurs at an intermediate critical temperature $T_c \approx 4.2$ (found by examining the absolute magnetization of sequences generated with only contacts, see Fig. \ref{fig:histo_magnetisation_graphs}, top panels). In the data set with only contacts, this phase transition strongly impacts partner inference performance, with a peak around $T_c$ apparent in Fig.~\ref{fig:partners_temp} (see also Refs.~\cite{Gandarilla20,NgampruetikornPreprint}). Qualitatively, at very low $T$, deep into the ferromagnetic phase, sequences are very similar to one another, as spins tend to all align, which makes inference difficult. At very high $T$, sequences become fully disordered and no longer reflect constraints from contacts, making inference difficult again. An optimum is thus expected at intermediate temperatures. The increased performance close to $T_c$ has been studied in detail in Ref.~\cite{NgampruetikornPreprint} for contact prediction, and no direct effects of criticality were found. Note that, more broadly, understanding which regime or phase is relevant in natural data is of high interest~\cite{Mora11,Morcos14}.

What is the impact of temperature when data are generated with both phylogeny and contacts? Fig~\ref{fig:partners_temp} shows that phylogeny substantially increases performance for $T>T_c$. More precisely, when phylogeny dominates ($\mu=15$, Fig~\ref{fig:partners_temp}A), partner inference performance is modest at low temperatures, but it improves as $T$ reaches $T_c$, and does not suffer the high-temperature decay observed in contact-only data when $T$ increases above $T_c$. Performance remains very good at large $T$, tending to the phylogeny-only performance value, which is high in this regime, consistently with Ref.~\cite{Marmier19}. We note that, in the low-temperature regime, performance is worse than with contact-only data. In the ordered phase, about half of the contact-only sequences include mainly $1$, while others include mainly $-1$ (see Fig.~\ref{fig:Magnetization}). In a typical species with four pairs AB comprising two sequences with mainly $1$ and two with mainly $-1$ (or three of one type and one of the other), pairing the A and B chains of the same overall sign is easy, but degeneracy makes distinguishing among them very hard, making the baseline expectation of partner performance about 50\% (more precisely, about 47\%, as there is a 1/8 probability to get all four sequences of the same type). By contrast, in data generated with contacts and phylogeny, evolution starts from an equilibrium ancestral chain AB, which is already mainly composed of either $1$ or $-1$. At low temperatures, switching the overall sign by successive mutations is difficult, and the whole phylogeny tends to retain the magnetization sign of its ancestor (see Fig.~\ref{fig:Magnetization}). Thus, the baseline expectation is only 25\% (random within-species matching), thereby explaining the lower performance of partner inference for data with contacts and phylogeny at low $T$. However, this ``freezing'' of states occurring in the low-$T$ limit in the present ferromagnetic Ising model is not expected to be the most relevant regime in real proteins. Thus, most other figures presented here correspond to $T=5>T_c$.

Let us now turn to the regime where phylogeny is less dominant ($\mu=30$, Fig~\ref{fig:partners_temp}B). While the low-temperature results are similar to those for $\mu=15$, performance now substantially decays in the high-temperature regime (this decay is very minor for $\mu=15$). Indeed, contacts become less informative as disorder increases, as for contact-only data. However, phylogeny makes this decay less strong, and at very high temperature, performance tends to the phylogeny-only value.  Fig~\ref{fig:partners_temp} also confirms that DCA and MI yield similar results, with MI becoming slightly better at high temperatures.

\paragraph{Impact of the training set size and of the number of pairs per species.} In addition to the parameters $\mu$ and $T$ that allow us to tune the relative importance of phylogeny and contacts, other parameters strongly impact inference. First, Fig.~\ref{fig:partners_trainingset} shows that a sufficiently large training set is required to accurately identify partners within each species. This holds both for DCA and for MI, but MI yields better performance for relatively small training sets, as for real data~\cite{Bitbol18}. While trends are similar for $\mu=15$ (Fig.~\ref{fig:partners_trainingset}A) and for $\mu=30$ (Fig.~\ref{fig:partners_trainingset}B), larger training sets are required to obtain the same performance in the latter case, confirming the positive impact of phylogeny on partner inference. The need for sufficiently large training sets also holds for data sets including only contacts and only phylogeny (see also Refs.~\cite{Marmier19,Gandarilla20}). These results are in line with previous ones obtained for DCA-based~\cite{Bitbol16,Gueudre16} and MI-based~\cite{Bitbol18} predictions of protein-protein interactions from natural protein sequence data. Second, the pairing task becomes more difficult when the number of pairs per species increases. Accordingly, Fig.~\ref{fig:NumberPairsSpecies} shows that the performance of partner inference decays as species contain more pairs AB. This decay is slowest for our data set including both contacts and phylogeny, highlighting that these two signals add constructively.

\paragraph{Impact of the graph of contacts.} How does the set of structural contacts impact the performance of partner inference? Because inter-protein contacts are generally sparser than intra-protein ones, we now consider graphs of contacts that take into account this constraint~\cite{Gandarilla20}, contrary to our minimal Erd\H{o}s-Rényi graph. Apart from the graph defining contacts, the data generation process and the inference procedures are exactly the same as before. Fig. \ref{fig:inference_temp_diff_graphs} shows that the same overall behavior is observed for the performance of partner inference for all graphs considered. With phylogeny, the range of temperature values leading to high performance is larger than for contact-only data sets. Moreover, for graphs possessing a smaller interface region between the two partners (Fig. \ref{fig:inference_temp_diff_graphs} C-D), the signal from contacts only does not suffice for good inference, and phylogeny then rescues inference. 

\subsection*{Couplings between non-contacting sites improve partner inference in the presence of strong phylogeny} 

\paragraph{Contributions of contacting and non-contacting sites to partner inference.} How does phylogeny improve the inference of partners? So far, we have shown that phylogeny often enhances the performance of partner inference. Indeed, partners share a common evolutionary history (here, they are generated together along the phylogenetic tree), and therefore, phylogeny yields correlations between sites that are informative of partnership. These phylogenetic correlations~\cite{Marmier19} are captured both by MI and DCA scores. Indeed, MI quantifies statistical dependence of any origin between random variables. The fact that DCA incorporates phylogenetic signal might seem more surprising, since it yields a Boltzmann distribution (Eq.~\ref{Boltz}) with a Potts Hamiltonian (Eq.~\ref{maxent}), thus formally resembling an equilibrium physical model. However, DCA approximately constructs the maximum-entropy distribution matching the one- and two-body frequencies measured in the training set. Its training objective is to match these empirical frequencies, whatever their origin. Thus, inferred DCA couplings incorporate phylogeny. In our minimal model, structural contacts only exist on the graph edges, and other (non-contact) pairs of sites have zero couplings in the Hamiltonian in Eq.~\ref{miniHam} used for data generation. Nonzero values of the inferred couplings between non-contact sites can arise due to phylogeny, but also due to finite-size effects, or to the approximations made in the inference procedure  (yielding for instance nonzero couplings for pairs of sites that are indirectly connected through other sites and thus correlated). Furthermore, the values of couplings between contacting sites can also be impacted by phylogeny, finite-size effects and inference approximations. How do couplings between contacting and non-contacting sites contribute to the inference in this synthetic data?

To address these questions, we restrict either to contact pairs of sites, or to non-contact pairs of sites in the score used for partner inference by DCA. Specifically, instead of the score $E_{AB}$ in Eq.~\ref{energy}, we use either
\begin{equation}
E^C_{AB}=-\sum_{i=1}^{L_A}\sum_{j=L_A+1}^{L_A+L_B} \hat{e}_{ij}(\alpha_i^A,\alpha_j^B)I_\mathcal{E}(i,j)\,,
\label{energyb}
\end{equation}
when restricting to contacts, or 
\begin{equation}
E^{NC}_{AB}=-\sum_{i=1}^{L_A}\sum_{j=L_A+1}^{L_A+L_B} \hat{e}_{ij}(\alpha_i^A,\alpha_j^B)\left[1-I_\mathcal{E}(i,j)\right]\,,
\label{energybb}
\end{equation}
when restricting to non-contacts, where $I_\mathcal{E}(i,j)$ is 1 if $(i,j)\in \mathcal{E}$ and 0 otherwise, $\mathcal{E}$ being the set of edges of the graph representing contacts. Because in our minimal model $e_{ij}(\alpha_i,\alpha_j)=\alpha_i\alpha_j$ if $(i,j)\in \mathcal{E}$ and 0 otherwise (see Methods), we also use the score
\begin{equation}
E^R_{AB}=-\sum_{i=1}^{L_A}\sum_{j=L_A+1}^{L_A+L_B} \alpha_i^A\alpha_j^B I_\mathcal{E}(i,j)\,,
\label{energyc}
\end{equation}
which would coincide with that in Eq.~\ref{energyb} if inference was perfect.

\begin{figure}[h]
	\centering
	\includegraphics[width=0.8\textwidth]{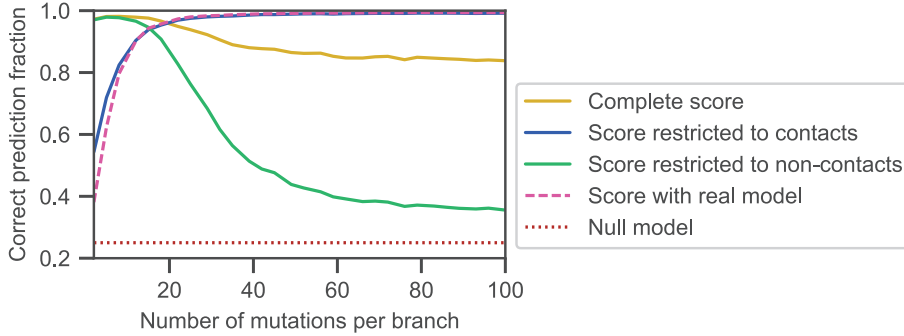}
	\caption{\textbf{Impact of contact and non-contact pairs of sites on partner inference performance.}  The fraction of correctly predicted partner pairs is shown versus the number $\mu$ of mutations per branch of the tree for the minimal model incorporating constraints both from contacts and from phylogeny, either with the full score defined in Eq.~\ref{energy} with couplings inferred using the training set, or with this score restricted to the pairs of sites that are actually in contact (Eq.~\ref{energyb}), or to those that are not in contact (Eq.~\ref{energybb}), or with a score computed with the real couplings used for data generation (Eq.~\ref{energyc}). Data generation and inference (apart from the score definition) are performed exactly as in Fig.~\ref{fig:partners_mutation}, using the same parameters and the same graph for contacts.}
	\label{fig:partners_noncontact}
\end{figure}

Fig.~\ref{fig:partners_noncontact} shows the performance of inference when these different scores are used on a data set containing constraints from contacts and phylogeny. When phylogeny dominates, i.e. for small $\mu$, restricting to contacts by using the score in Eq.~\ref{energyb} strongly deteriorates the performance of inference. Thus, couplings between non-contacting sites then include phylogenetic information relevant to infer partners. Conversely, when the effect of phylogeny is smaller, i.e. for large $\mu$, restricting to contacts improves inference performance compared to using the full score. This deleterious impact of non-contacting pairs is probably due to the fact that these couplings also arise from finite-size noise and inference approximations, in addition to phylogeny. Results obtained from the score based on the real Hamiltonian Eq.~\ref{energyc} are close to those obtained with Eq.~\ref{energyb}, and even a little worse when phylogeny is very strong, because the real Hamiltonian cannot capture any phylogeny.

By contrast, restricting to non-contact pairs via the score in Eq.~\ref{energybb} yields an inference performance as good as with the complete score in Eq.~\ref{energyb} for very small $\mu$, confirming that non-contact pairs incorporate most phylogenetic signal. As $\mu$ increases and phylogenetic signal weakens, performance using the score in Eq.~\ref{energybb} decays sharply. This decay resembles that observed in Fig.~\ref{fig:partners_mutation} for partner inference on purely phylogenetic data, confirming that non-contact couplings mainly arise from phylogeny. However, performance using the score restricted to non-contact pairs remains slightly higher than the null model for large $\mu$, while it tends to the null model with purely phylogenetic data. Contrary to most of the signal associated to non-contact sites, this small residual performance cannot be attributed to phylogeny, and might arise from the fact that non-contact couplings contain indirect correlations, stemming from contacts but mediated by intermediate sites. This is despite the fact that DCA reduces such contributions compared to covariance or mutual information~\cite{Weigt09,NgampruetikornPreprint}. The importance of indirect correlations could be impacted by the approximations made when inferring couplings. To explore this, we employed couplings inferred by bmDCA, which yields a much better inference quality than mfDCA in terms of generative properties~\cite{Figliuzzi18}, but is computationally much heavier. Fig.~\ref{fig:partners_noncontact_bm} shows that the results using bmDCA are almost identical to those obtained by mfDCA, which shows the robustness of our conclusions. Note that even sites that are not coupled to any other ones carry some phylogenetic signal, as illustrated by Fig.~\ref{fig:partners_single_interf}.

\newpage

\begin{figure}[h]
	\centering
	\includegraphics[width=\textwidth]{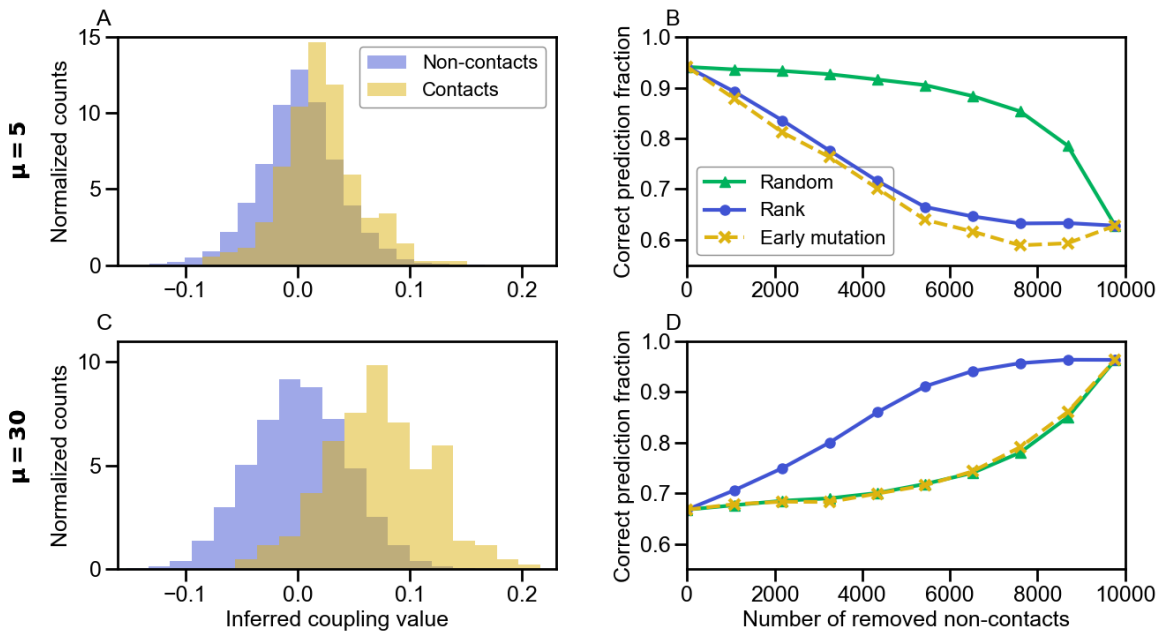}
	\vspace{0.01cm}
	\caption{\textbf{Impact of gradually removing non-contact pairs of sites on partner inference performance.} Panels A and C show the normalized histogram of the inferred couplings $\hat{J}_{ij}$ for contact and non-contact pairs of sites. Panels B and D show the fraction of correctly predicted partner pairs versus the number of removed non-contact pairs of sites, when gradually removing their contribution from the score in Eq.~\ref{energy}, going toward the one in Eq.~\ref{energyb}. The non-contact pairs are removed either randomly, or in decreasing rank order of $|\hat{J}_{ij}|$ values, or in increasing order of the earliest time (i.e., earliest generation in the tree) when mutations have affected both sites $i$ and $j$, not necessarily in the same sequence. In A and B, $\mu=5$, while in C and D, $\mu=30$. All data is shown for the minimal model incorporating both constraints from contacts and phylogeny. Data generation and inference (apart from the score definition) are performed exactly as in Fig.~\ref{fig:partners_mutation}, using the same graph for contacts and the same parameters, except that here, the training set comprises 100 paired chains AB. In panels A and C, data is generated just once. In panels B and D, each point is averaged over 100 generations of the data. }
	\label{fig:coupl}
\end{figure}

\paragraph{Impact of gradually removing non-contacting pairs of sites.} To gain further insight into the impact of non-contact pairs of sites on partner inference, we next investigate the impact of removing them gradually. We remove them either randomly, or by decreasing rank order of the absolute value of the inferred couplings, or based on mutation timing. Indeed, non-contact pairs may have high impact because they have large absolute inferred coupling values, and/or because they feature strong phylogenetic correlations. Fig.~\ref{fig:coupl} shows histograms of the values of inferred Ising couplings $\hat{J}_{ij}$, defined by $\hat{e}_{ij}(\alpha_i,\alpha_j)=\hat{J}_{ij}\alpha_i\alpha_j$, between contact and non-contact pairs of sites, for $\mu=5$ (Fig.~\ref{fig:coupl}A) and for $\mu=30$ (Fig.~\ref{fig:coupl}C). In the first case, where phylogeny dominates, the values of $\hat{J}_{ij}$ do not allow one to distinguish contacts from non-contacts. In the second case, contact pairs feature higher values of $\hat{J}_{ij}$, which would allow to infer them better, although there is still a strong overlap, partly due to the rather small training set (100 sequences) employed here, which was chosen for the partner inference task to be successful with the score in Eq.~\ref{energy}, but still improvable (see Fig.~\ref{fig:partners_trainingset}). Next, Fig.~\ref{fig:coupl} B and D show the impact of progressively removing non-contact couplings on the performance of partner inference. For small $\mu$, removing them decreases performance, while the opposite holds for large $\mu$, consistent with Fig.~\ref{fig:partners_noncontact}. Moreover, for $\mu=5$ (Fig.~\ref{fig:coupl}B), removing them in decreasing rank order of absolute $\hat{J}_{ij}$ yields a quicker and sharper decay of performance than removing them in random order. Thus, large absolute non-contact couplings contain relevant information for partner inference. Removing them in increasing order of the earliest time along the phylogeny when mutations have affected both sites $i$ and $j$, not necessarily in the same sequence, yields a similarly quick and sharp decay, corroborating the idea that phylogeny is the main relevant source of information in these non-contacting pairs. Indeed, early coupled mutations lead to large phylogenetic correlations. Conversely, for $\mu=30$ (Fig.~\ref{fig:coupl}D), ranking-based removal of non-contacts leads to an earlier increase of performance than random removal, but earliest-mutation-based removal is almost equivalent to random removal. This illustrates the reduced importance of phylogeny in this case.

\newpage

\subsubsection*{Generalization to another phylogeny model}

So far, we have considered a model where sequences generated along a phylogeny are randomly grouped into ``species". Such random species are realistic if exchanges of genes between species (i.e. horizontal gene transfer events) are very frequent. But more generally, different levels of relatedness are expected between sequences within and across species. To assess the robustness of our results to this point, we also considered another, completely different, phylogeny model, with duplication, loss and speciation events, and without any exchange between species~\cite{Marmier19}. In this model, the number $m$ of paired sequences per species is fixed.

Fig.~\ref{fig:arbre_alternatif} shows the performance of inference versus the fraction of species that undergo a duplication-loss event upon speciation. Overall, performance is good, but it decreases when the frequency of duplication-loss events increases, especially if there is substantial signal from phylogeny (small to intermediate $\mu$). Indeed, when there are no duplication-losses, this model features a distinct phylogeny for each ancestral sequence, i.e. $m$ different phylogenies ($m=4$ in Fig.~\ref{fig:arbre_alternatif}). Similarities between the chains of the testing and training set that share the same ancestor then allow to predict evolutionary partners, and contacts are useful too, yielding very good inference performance. Conversely, when there are many duplication-losses, chains descending from a single ancestral protein will take over in each species, analogously to the mutant fixation process in population genetic models such as the Moran model~\cite{Ewens79}. Thus, phylogenetic correlations gradually become less useful, and even hurtful compared to the contact-only case, when the frequency of duplication-loss events increases. Moreover, chains resulting from recent duplication-loss events are very hard to distinguish, making inference difficult in this regime~\cite{Marmier19}. Fig.~\ref{fig:arbre_alternatif}A further shows that when the probability of duplication-loss is small, correlations from contacts and phylogeny add constructively. This is consistent with the results obtained above with our first phylogeny model. Fig.~\ref{fig:arbre_alternatif}B shows the impact of restricting to contacts or to non-contacts in the pairing score used for partner inference. Restricting to contacts yields better results than restricting to non-contacts for large enough values of $\mu$, but the opposite holds for small ones ($\mu=5$) where phylogenetic correlations play a crucial role. This is consistent with our other results (see Fig.~\ref{fig:partners_noncontact}). Hence, our main conclusions hold for two very different phylogeny models, demonstrating their robustness.

%\newpage

\subsection*{Interplay of contacts and phylogeny in natural data and in synthetic data generated from models inferred on natural data}

Our minimal model allows us to tune the importance of contacts and phylogeny, but contains strong simplifications. Natural data comprises $q=21$ possible states, which are the 20 natural amino acids and the alignment gap. Potts models inferred from natural data involve broad distributions of couplings, and include nonzero fields~\cite{Weigt09,Marks11,Morcos11,RodriguezHorta21}. Phylogenies inferred from data are also much more complex than our binary tree with a fixed number of mutations on each branch~\cite{Price10,Vorberg18}, and the assignment of sequences to species results from speciation and horizontal gene transfer. Do the conclusions obtained with our minimal model hold for natural data? How important are signals from phylogeny and contacts in natural data? To address this question, we consider two data sets of interacting pairs of natural sequences. The first one is composed of 23,633 pairs of histidine kinases (HK) and response regulators (RR) from the P2CS database~\cite{Barakat09,Barakat11}. The second one comprises 17,950 pairs of ABC transporter proteins homologous to the \textit{Escherichia coli} MALG-MALK pair of maltose and maltodextrin transporters~\cite{Ovchinnikov14,Bitbol16}. To gain further insight into the importance of phylogenetic signal, we infer generative models of this paired sequence alignment using two state-of-the-art methods, bmDCA \cite{Figliuzzi18} and arDCA \cite{Trinquier21} (see Methods), and we generate data from them, either without phylogeny or along a phylogenetic tree inferred from the natural alignment~\cite{Vorberg18}. Importantly, in the latter case, we retain the species labels of the leaves of the inferred tree and we use them to group synthetic sequences into species (see Methods).  Hence, the relationship between species and positions on the tree is the same in this synthetic data generated with phylogeny as in the natural data. Note that we employed both bmDCA and arDCA for the HK-RR data set, but only arDCA for the ABC transporter data set, because the longer sequences yield very long computation times with bmDCA in that case. We checked that two-body and one-body frequencies of the original data set were well-reproduced by those of the data set generated without phylogeny (Fig. \ref{fig:two-body-freq}). We also checked that the inference of contacts was possible on the generated data sets, although it deteriorates when generating with phylogeny (Fig. \ref{fig:21_inference_contact}).

\begin{figure}[h]
	\centering
	\includegraphics[width=0.9\textwidth]{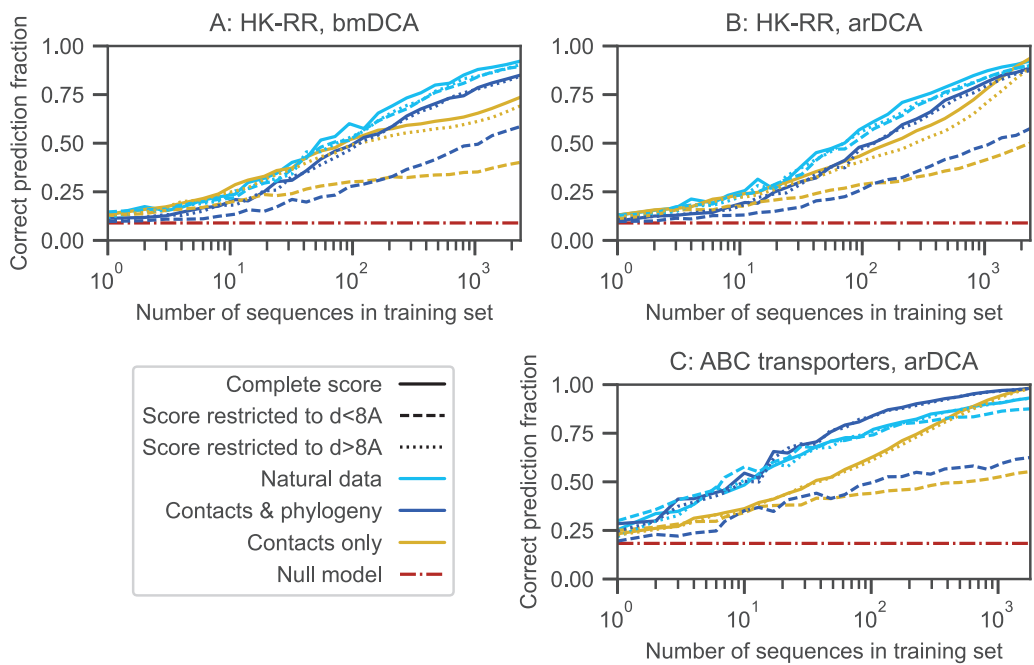}
	\caption{\textbf{Partner inference in natural data and in synthetic data generated from models inferred on natural data.} The fraction of correctly predicted partner pairs is shown versus the number of sequence pairs in the training set for two different natural data sets of paired sequences, HK-RR (panels A and B) and ABC transporters (panel C). For both of them, we also present results obtained on synthetic data sets generated using models learned from the corresponding natural data set by bmDCA (panel A) or arDCA (panels B and C), either with or without phylogeny. Recall that pairings are predicted within each species. For this, species identifiers are obtained for natural sequences. When generating synthetic data along inferred phylogenetic trees, we retain the species identifier of the original natural sequence corresponding to this leaf, and attribute this species identifier to the synthetic sequence that is generated on this leaf. This yields artificial data sets with a similar species structure as the natural ones. In each case, the full data set is split into a training set and a testing set, each comprising a given number of species (and thus of paired sequences). Partnerships are then predicted using the score in Eq.~\ref{energy} based on mfDCA, or with its variants restricting to amino-acid pairs in contact (Eq.~\ref{energyb}) or to those not in contact (Eq.~\ref{energybb}). Contacts are defined using a threshold of 8 \AA~between closest atoms in the experimental complex structures 3DGE for HK-RRs and 3RLF for ABC transporters, retrieved from the PDB~\cite{pdb}. The null model shows the expectation of the correct prediction fraction if pairs are made randomly within each species.}
	\label{fig:21_inferecene_partner}
\end{figure}

Fig. \ref{fig:21_inferecene_partner} shows the fraction of correct predicted pairs versus the size of the training set for the natural data set, as well as for the synthetic data sets generated from inferred models. Both for HK-RRs and ABC transporters, results employing the usual mfDCA-based score in Eq.~\ref{energy} are qualitatively similar for the real and synthetic data sets, as well as to the results from our minimal model (Fig.~\ref{fig:partners_trainingset}). Furthermore, the performance of partner inference is similar for synthetic data sets generated from inferred models with and without phylogeny. This is \textit{a priori} reminiscent of cases with relatively low phylogeny in the minimal model (Fig.~\ref{fig:partners_trainingset}B). However, DCA models inferred from natural data reproduce all empirical correlations, including those from phylogeny. Some inferred couplings are thus of phylogenetic origin, and give rise to correlations in the sequences generated from these models without phylogeny. These additional couplings can help partner inference, by extending the set of pairs of sites that can yield information relevant for pairing, and by increasing the diversity of covariance and coupling values. This may contribute to the performance of partner inference in the synthetic data set generated without phylogeny. This effect can be assessed in our minimal model, by generating data from models inferred by bmDCA or arDCA (the initial data employed for inference being generated using the Hamiltonian in Eq.~\ref{miniHam}, with or without phylogeny, see Figs.~\ref{fig:fij1} and~\ref{fig:fij2}). Fig.~\ref{fig:partners_inferc} shows that in our minimal model, partner inference performance is higher with data generated from inferred models, compared to similar data generated directly from the original Hamiltonian in Eq.~\ref{miniHam}. This demonstrates that inferred models incorporating couplings from various sources, including but not restricted to phylogeny, yield better partner inference performance than contact-only models. This effect explains at least partly the performance of partner inference for synthetic data sets generated from more realistic inferred models but without phylogeny in Fig.~\ref{fig:21_inferecene_partner}.

In this context, it is interesting to investigate the effect of restricting to contact pairs of amino-acids or to non-contact ones. Fig. \ref{fig:21_inferecene_partner} shows partner inference performance for the reduced scores in Eqs.~\ref{energyb} and~\ref{energybb}. It demonstrates that restricting to contact pairs (Eq.~\ref{energyb}) tends to deteriorate partner inference performance, to a minor extent for natural data (even using the generous distance threshold of 8~\AA~between closest atoms to define contacts -- but note that the contacts observed in crystal structures are not necessarily all those that exist \textit{in vivo}), and more strongly for data generated using DCA models inferred from natural data. This difference is consistent with the less good contact inference from generated data, see Fig. \ref{fig:21_inference_contact}. Moreover, restricting to non-contact pairs (Eq.~\ref{energybb}) only yields a minor decrease of inference performance for all these data sets, compared to using the full score (Eq.~\ref{energy}). Thus, strikingly, the information contained in non-contact pairs is sufficient for inference. Overall, our results on natural data and on realistic synthetic data are consistent with what is observed for intermediate to relatively-strong phylogeny in our minimal model (Figs.~\ref{fig:partners_noncontact}, \ref{fig:coupl} and \ref{fig:partners_inferc}).

%\clearpage

\section*{Discussion}

While they obscure the identification of contacts by coevolution methods~\cite{Dunn08,Weigt09,Marks11,Qin18,RodriguezHorta21}, correlations that arise in protein sequences due to phylogeny~\cite{Casari95,Halabi09,Qin18,Fryxell96,Goh00,Pazos01,Ochoa10} become useful in order to identify interaction partners from protein sequences. Indeed, interaction partners tend to have similar evolutionary histories, which is directly exploited in some protein-protein interaction prediction methods~\cite{Pazos01,Jothi05,Bradde10,Ochoa10,Ochoa15,Pellegrini99,Croce19,Moi20}. In this context, the success of DCA- and MI-based approaches at predicting protein-protein interaction partners among paralogs from natural protein sequences~\cite{Bitbol16,Gueudre16,Bitbol18} can potentially be due to correlations from structural contacts needing to maintain their complementarity, or to correlations from phylogeny, or both. Shedding light on the origin of the performance of these methods is an important step toward constructing better ones. However, disentangling the impact of different sources of signal in natural data is a difficult task.

In this study, we generated and analyzed synthetic data produced within a minimal model that allows us to control the amounts of structural constraints and phylogeny. We showed that these two signals add constructively to increase the performance of inference of partners among paralogs by DCA or MI. Furthermore, signal from phylogeny can rescue partner inference in cases where signal from structural contacts becomes less informative, including in the realistic case where inter-protein contacts are restricted to a small subset of sites. We also demonstrated that DCA-inferred couplings between non-contact pairs of sites improve partner inference in the presence of strong phylogeny, while deteriorating it otherwise. Furthermore, in the strongly phylogenetic regime, inference is almost as good when restricting to non-contact pairs of sites than when including all pairs. An important advantage of our controlled synthetic data is that functional constraints are exactly known because they are an input of the model (the couplings in the Hamiltonian). Moreover, tuning the strength of functional and phylogenetic signals allowed us to determine their respective roles, while disentangling them in natural data is difficult. However, our minimal model is a strong simplification compared to natural data. To make the connection, we next considered a natural data set, as well as realistic synthetic data based on it. We confirmed that non-contact pairs of sites contribute positively to partner inference performance. Moreover, restricting to non-contact pairs of amino acids yields inference performances that are very close to those obtained when all pairs are accounted for. These results are in line with what was observed in our minimal model with strong phylogeny, and evidence an important role of phylogeny in partner inference on natural data.

While we demonstrated that phylogenetic correlations are helpful to determine partners among the paralogs of two protein families, the importance of their contribution is an issue for the task of determining whether two protein families interact or not. Indeed, while families of proteins in direct physical interaction often have similar evolutionary histories~\cite{Pazos01,Jothi05,Bradde10,Ochoa10,Ochoa15}, this can also be the case for proteins that are not directly interacting, e.g. those that belong to the same pathway, or are expressed in the same operon in bacteria, or for proteins with similar evolutionary rates~\cite{Hakes07,Juan08,Kann09,Lovell10,Swapna12}. Nevertheless, physically interacting protein families can be distinguished from others using the top DCA scores between amino acids~\cite{Marmier19,Cong19,Green21}, or alternatively, the top predicted contact probability over all amino acid pairs~\cite{Humphreys21} or a docking quality score~\cite{EvansPreprint,Bryant22} in deep learning approaches. Importantly, all these methods, whether they are DCA-based or deep learning-based, start from a paired multiple sequence alignment of homologs of the two candidate interaction partners. The methods discussed here, which allow to predict partners among the paralogs of two protein families, can potentially improve the depth and quality of these paired alignments.

Several interesting extensions are possible. First, in this work, we considered the problem of partner inference starting from a training set of known partners, but our methods allow us to address the case where there is no training set via an Iterative Pairing Algorithm (IPA)~\cite{Bitbol16, Bitbol18}. It would be interesting to extend the present study to this case. Next, we assumed for simplicity that there was a strict one-to-one pairing between partners. Natural systems can be more complex, e.g. involving different numbers of members of the two families considered within each species, or proteins with multiple partners (i.e., promiscuity or cross-talk). The first point can be addressed by a slight modification of our approach, using an ``injective matching strategy''~\cite{Gueudre16} that leaves unpaired some sequences of the family with more paralogs. The second point could be tackled through variants of our approach and of the IPA~\cite{Bitbol16}, e.g. by introducing a threshold on the pairing score, and retaining all candidate pairs passing this threshold, even if several involve a given sequence. In practice, knowing that interactions are one-to-one in some model species could be used to make one-to-one predictions in other species, but generalizing the IPA to allow for multiple partners is an important direction for future work. Besides, while our study of synthetic data generated using models inferred from natural data allowed us to bridge our minimal model and natural data  and to assess the importance of signal from contacts and from phylogeny in these more realistic cases, we were faced with the issue that inferred couplings include phylogeny. Thus, disentangling signals was much harder than in the minimal model, as the couplings from phylogeny make the model richer even in the absence of phylogeny in the data generation step. While this is a difficult problem, it could be partially addressed by applying phylogeny corrections to the inferred couplings~\cite{RodriguezHorta21,Colavin22}. This could also shed light on whether some of the useful signal from non-contact pairs is coming from collective functional constraints, similar to sectors in single proteins~\cite{Halabi09,Wang19,Colavin22}, an interesting possibility that was not explored here. Investigating the impact of imperfections in protein sequence alignment on partner inference would also be highly relevant, as well as including the possibility of one-to-many pairings and crosstalk. Finally, understanding the relative impact of structural constraints and phylogeny in the inference of interaction partners from sequences opens the way to exploiting them together more efficiently.

\section*{Acknowledgments}

This project has received funding from the European Research Council (ERC) under the European Union’s Horizon 2020 research and innovation programme (grant agreement No. 851173, to A.-F. B.).

%\nolinenumbers

% Either type in your references using
% \begin{thebibliography}{}
% Text
% \end{thebibliography}
%
% or
%
% Compile your BiBTeX database using our plos2015.bst
% style file and paste the contents of your .bbl file
% here. See http://journals.plos.org/plosone/s/latex for 

\newpage

\section*{Supporting information}
\def\theequation{S\arabic{equation}}
\setcounter{equation}{0}
\def\thefigure{S\arabic{figure}}
\def\thetable{S\arabic{table}}
\setcounter{figure}{0}

% Include only the SI item label in the paragraph heading. Use the \nameref{label} command to cite SI items in the text.

\begin{figure}[h]
	\centering
	\includegraphics[width=0.9\textwidth]{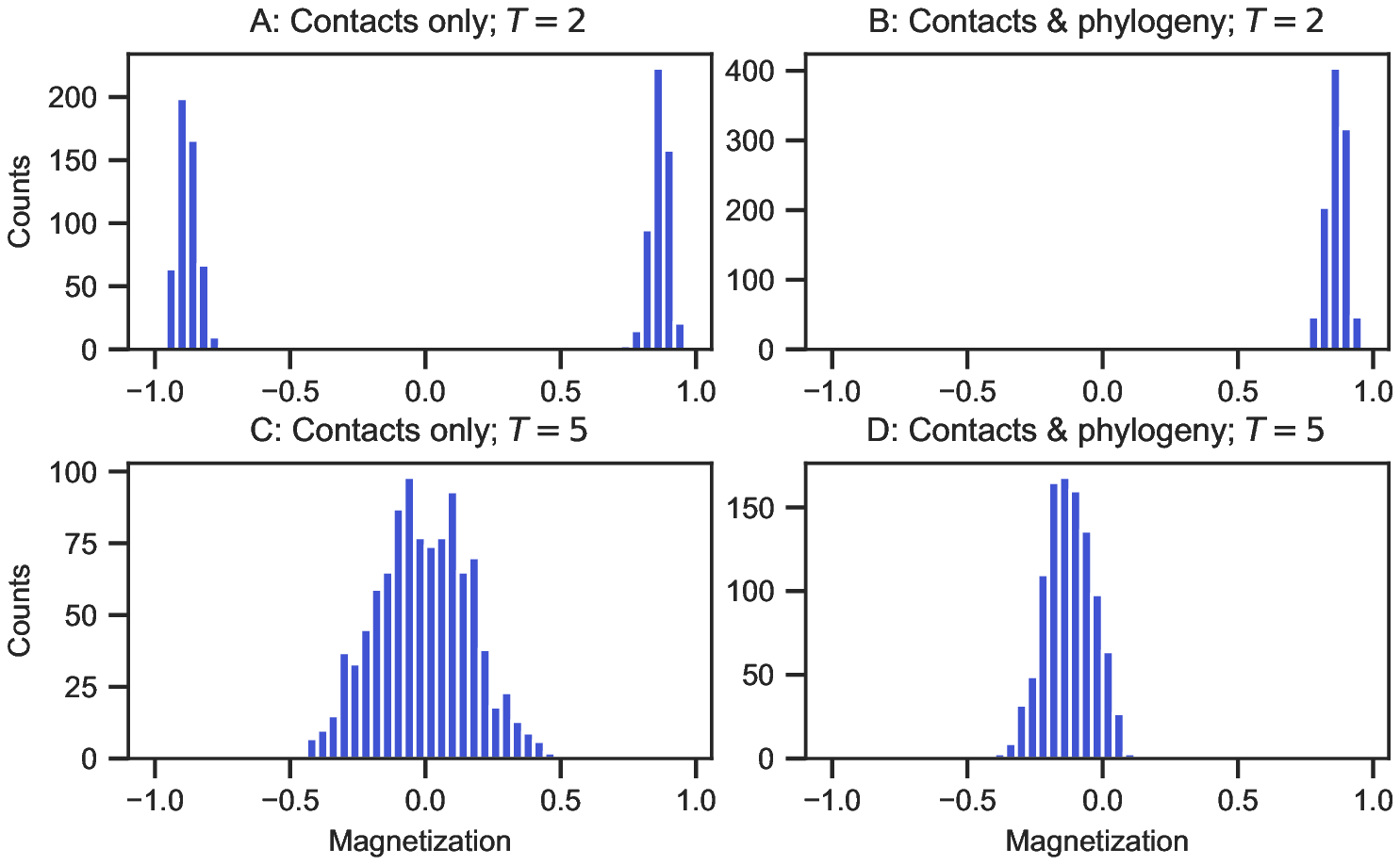}
	\caption{\textbf{Magnetization for data generated without and with phylogeny.} Histograms of the mean magnetization $m=(\sum_{i=1}^{2L}\alpha_i)/(2L)$ of a paired chain are shown at two different sampling temperatures $T$ ($T=2<T_c$ in panels A and B; $T=5>T_c$ in panels C and D) for data generated without (contacts only, panels A and C) or with phylogeny (contacts \& phylogeny, panels B and D). Contacts are defined by the same Erd\H{o}s-Rényi graph with $p=0.02$ as in Fig.~\ref{fig:partners_mutation}. In the contact-only case, histograms are computed on an equilibrium data set of $1024$ independent sequences generated by Metropolis Monte Carlo sampling under the Hamiltonian in Eq.~\ref{miniHam} (without phylogeny). In practice, data is taken after 10,000 accepted mutations, see Fig.~\ref{fig:magnetisation_against_mutations}. In the contacts \& phylogeny case, histograms are computed on a data set of $1024$ sequences generated on a single binary branching tree with $\mu=15$ mutations per branch.}
	\label{fig:Magnetization}
\end{figure}

\begin{figure}[h!]
	\centering
	\includegraphics[width=0.9\textwidth]{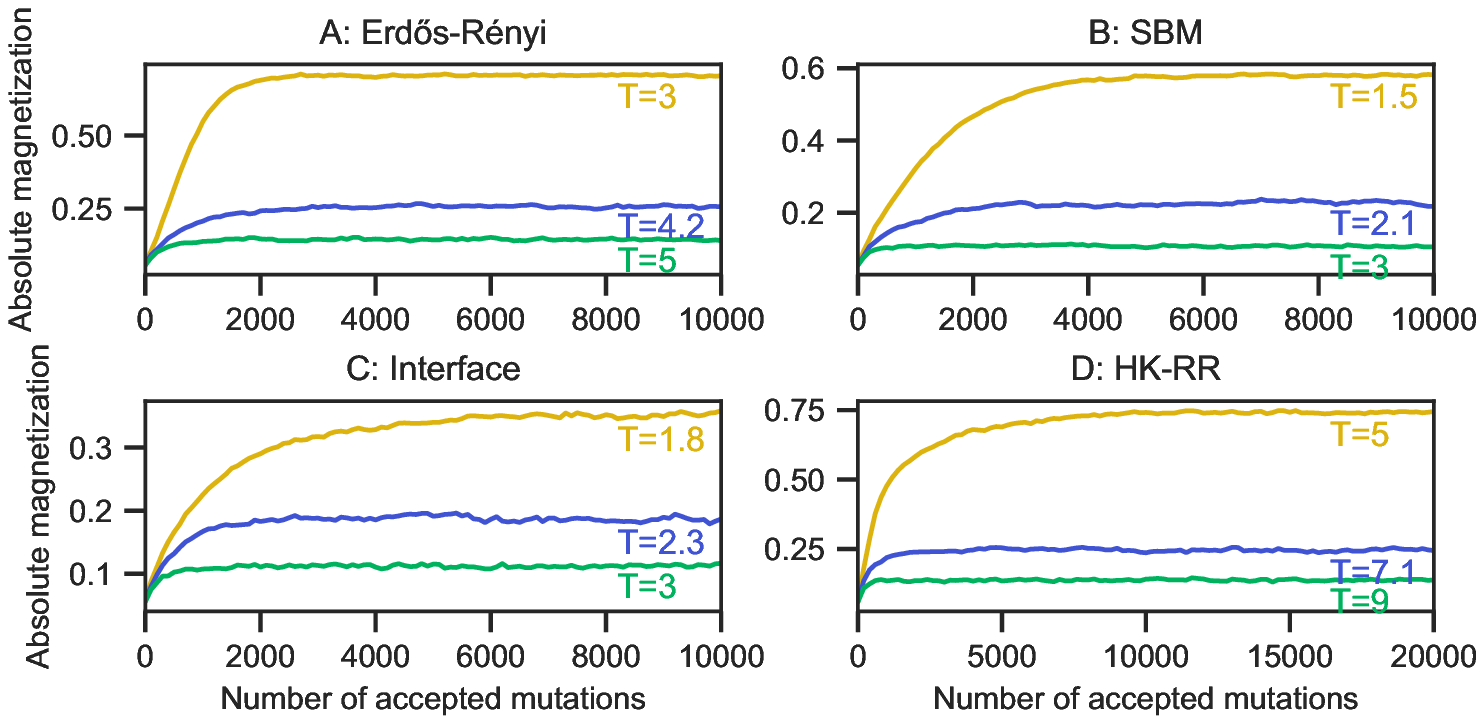}
	\caption{\textbf{Convergence of absolute magnetization in the contact-only limit with different graphs.} The mean of the absolute magnetization $|m|=|(\sum_{i=1}^{2L}\alpha_i)/(2L)|$ of a paired chain is shown versus the number of accepted mutations for each graph considered in Fig. \ref{fig:inference_temp_diff_graphs}. For each curve, data sets of $5000$ independent sequences were generated by Metropolis Monte Carlo sampling under the Hamiltonian in Eq.~\ref{miniHam} (without phylogeny). The graphs considered are: (A) the same Erd\H{o}s-Rényi graph with $p=0.02$ as in Fig.~\ref{fig:partners_mutation}; (B) a stochastic block model graph with two blocks of 100 nodes each, and $p=0.02$ within each block and $p=0.005$ between blocks; (C) an ``interface'' graph with two blocks of 100 nodes each, and $p=0.02$ within each block, but where only 10 nodes in each block are allowed to be in contact with nodes of the other block, with $p=0.25$; (D) a graph corresponding to the contact map from the experimental HK-RR complex structure in Ref.~\cite{Casino09} with threshold at 4 \AA~between closest atoms.}
	\label{fig:magnetisation_against_mutations}
\end{figure}

\begin{figure}[h!]
	\centering
	\includegraphics[width=\textwidth]{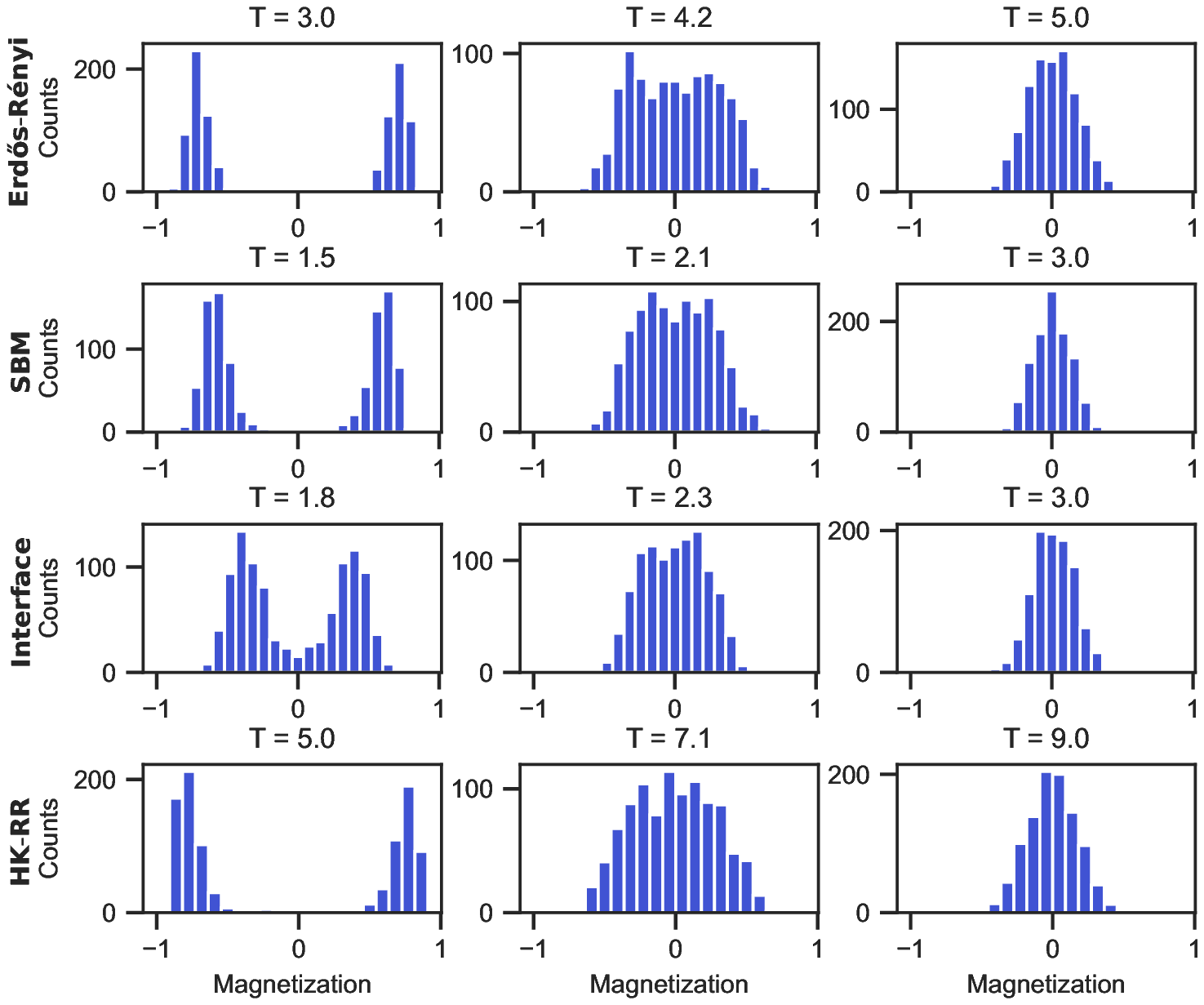}
	\caption{\textbf{Ferromagnetic-paramagnetic phase transitions with different graphs.} Histograms of the mean magnetization $m=(\sum_{i=1}^{2L}\alpha_i)/(2L)$ of a paired chain are shown at different sampling temperatures $T$ for each graph considered in Fig. \ref{fig:inference_temp_diff_graphs}. Histograms are computed on 5000 independent sequences generated by Metropolis Monte Carlo sampling under the Hamiltonian in Eq.~\ref{miniHam} (without phylogeny). In practice, data is taken after 10,000 accepted mutations (20,000 for the HK-RR graph), see Fig.~\ref{fig:magnetisation_against_mutations}. The graphs considered are: (A) the same Erd\H{o}s-Rényi graph with $p=0.02$ as in Fig.~\ref{fig:partners_mutation}; (B) a stochastic block model graph with two blocks of 100 nodes each, and $p=0.02$ within each block and $p=0.005$ between blocks; (C) an ``interface'' graph with two blocks of 100 nodes each, and $p=0.02$ within each block, but where only 10 nodes in each block are allowed to be in contact with nodes of the other block, with $p=0.25$; (D) a graph corresponding to the contact map from the experimental HK-RR complex structure in Ref.~\cite{Casino09} with threshold at 4 \AA~between closest atoms. These graphs are exactly the same as in Fig.~\ref{fig:magnetisation_against_mutations}.}
	\label{fig:histo_magnetisation_graphs}
\end{figure}

\begin{figure}[h]
	\centering
	\includegraphics[width=\textwidth]{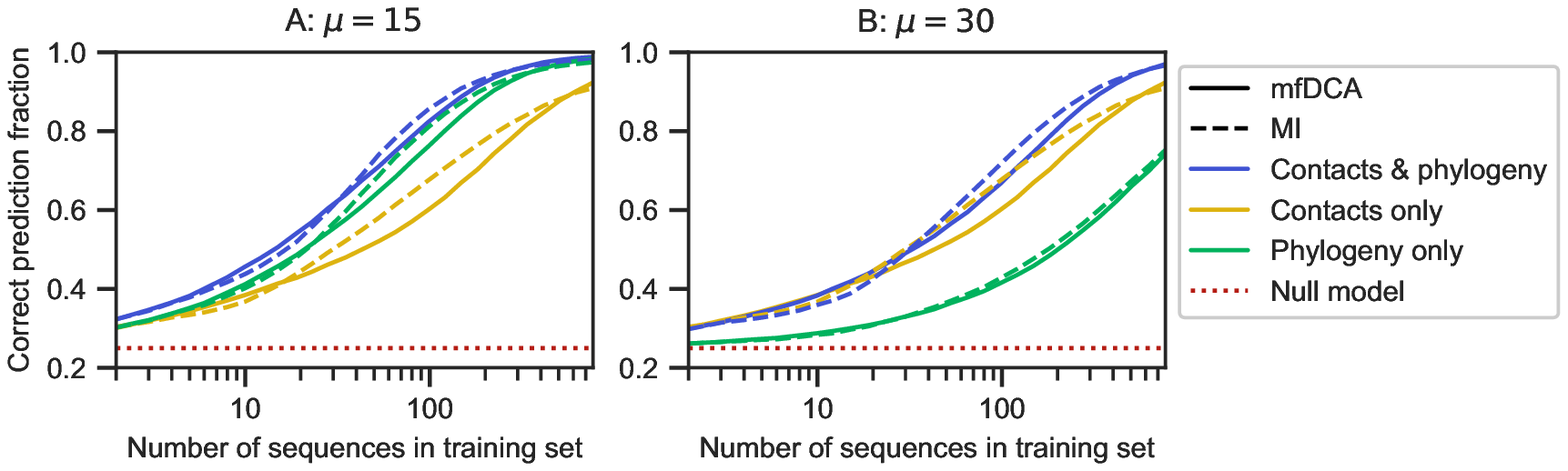}
	\caption{\textbf{Impact of training set size on partner inference performance.} The fraction of correctly predicted partner pairs is shown versus the number of sequence pairs AB in the training set for the minimal model incorporating both constraints from contacts and phylogeny, exactly as in Fig.~\ref{fig:partners_mutation}. Specifically, an ancestral chain AB of $2L=200$ spins is evolved along a binary branching tree (see Fig. \ref{fig:phylo_tree}) with $n=10$ generations, and $\mu=15$ (panel A) or $\mu=30$ (panel B) mutations per branch, yielding $2^{10}=1024$ pairs AB, at sampling temperature $T=5$ under the Hamiltonian in Eq.~\ref{miniHam} on the same graph as in Fig.~\ref{fig:partners_mutation}. The limiting cases with only contacts and only phylogeny are also shown for comparison. The first one corresponds to independent equilibrium sequences at $T=5$ under the Hamiltonian in Eq.~\ref{miniHam} on the same graph. The second one corresponds to neutral evolution on the same binary branching tree. Each of these three data sets is randomly split into a training set comprising a variable number of chains AB and a testing set. The latter is randomly divided in species of 4 chains AB each, and pairings between A and B chains are blinded in each species. Partnerships are then predicted using the scores in Eq.~\ref{energy} (mfDCA) or Eq.~\ref{SAB} (MI). Each point is averaged over 20 generated data sets, and 20 random choices of the training set for each of them. The null model shows the expectation of the correct prediction fraction if pairs are made randomly within each species.}
	\label{fig:partners_trainingset}
\end{figure}

\begin{figure}[h]
	\centering
	\includegraphics[width=\textwidth]{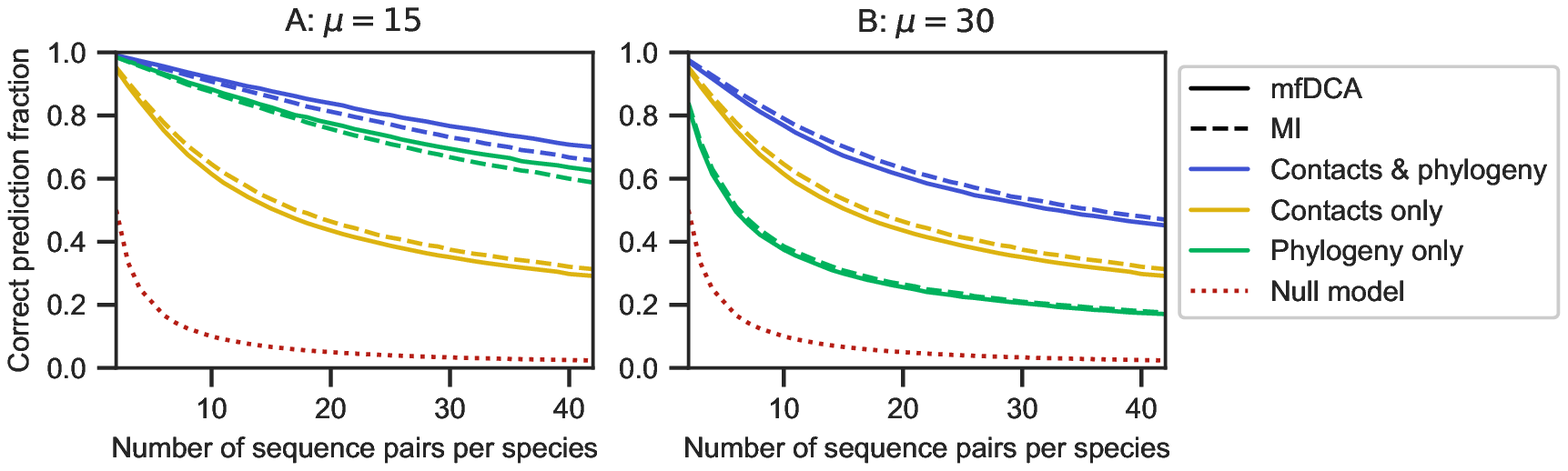}
	\caption{\textbf{Impact of the number of pairs per species on partner inference performance.} The fraction of correctly predicted partner pairs is shown versus the number $m$ of sequence pairs AB per species in the testing set for the minimal model incorporating both constraints from contacts and phylogeny, exactly as in Fig.~\ref{fig:partners_mutation}. Specifically, an ancestral chain AB of $2L=200$ spins is evolved along a binary branching tree (see Fig. \ref{fig:phylo_tree}) with $n=10$ generations, and $\mu=15$ (panel A) or $\mu=30$ (panel B) mutations per branch, yielding $2^{10}=1024$ pairs AB, at sampling temperature $T=5$ under the Hamiltonian in Eq.~\ref{miniHam} on the same graph as in Fig.~\ref{fig:partners_mutation}. The limiting cases with only contacts and only phylogeny are also shown for comparison. The first one corresponds to independent equilibrium sequences at $T=5$ under the Hamiltonian in Eq.~\ref{miniHam} on the same graph. The second one corresponds to neutral evolution on the same binary branching tree. Each of these three data sets is randomly split into a training set of 400 chains AB and a testing set of 624 chains AB. The latter are randomly divided in species of $m$ chains AB each (here $m$ is varied) and pairings between A and B chains are blinded in each species. Partnerships are then predicted using the scores in Eq.~\ref{energy} (mfDCA) or Eq.~\ref{SAB} (MI). Each point is averaged over 20 generated data sets, and 20 random choices of the training set for each of them. The null model shows the expectation of the correct prediction fraction if pairs are made randomly within each species.}
	\label{fig:NumberPairsSpecies}
\end{figure}

\begin{figure}
	\centering
	\includegraphics[width=0.9\textwidth]{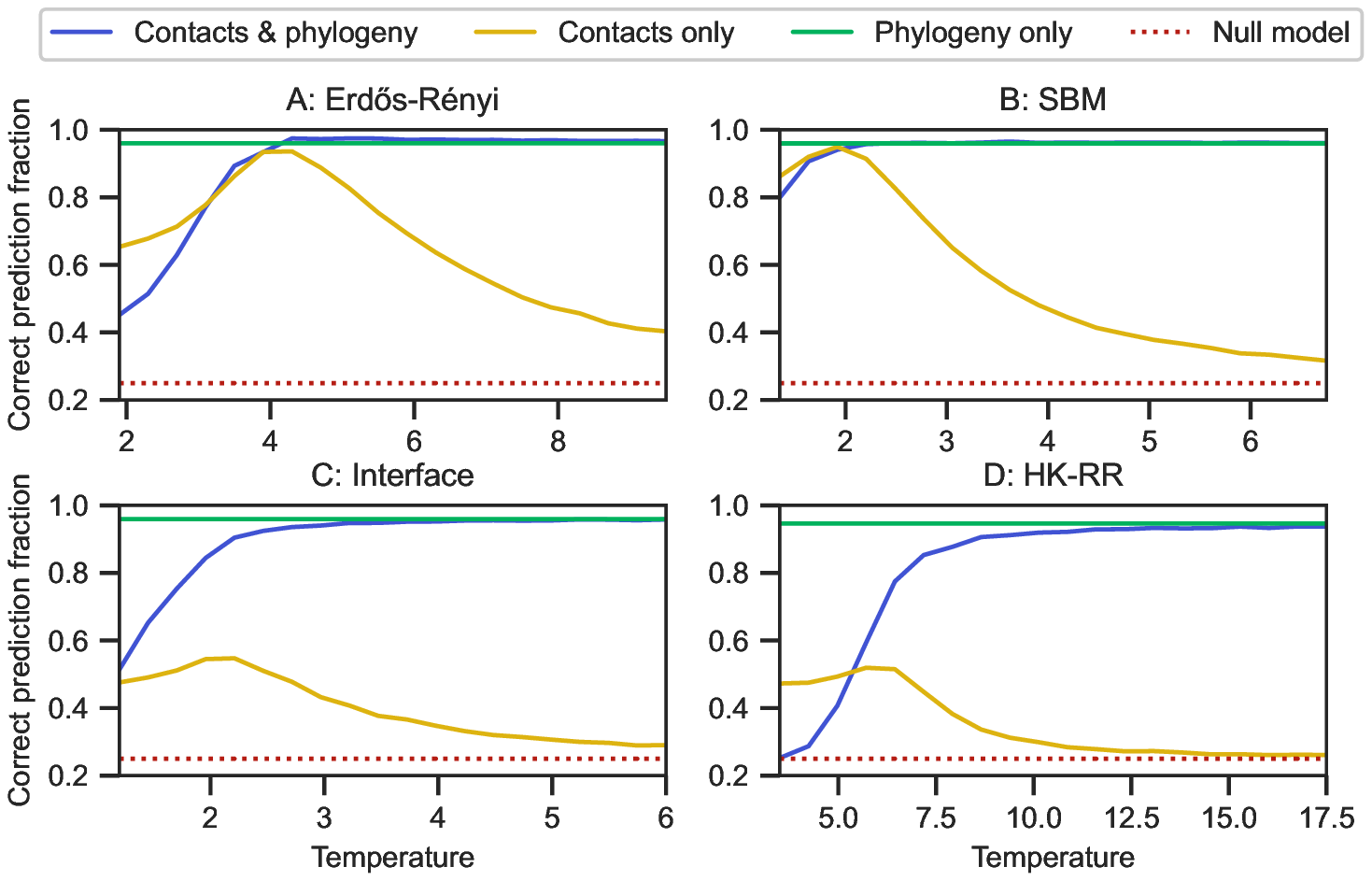}
	\caption{\textbf{Partner inference performance for different graphs defining contacts.} For four different graphs (panels A to D), the fraction of correctly predicted partner pairs is shown versus the sampling temperature $T$ for the minimal model incorporating both constraints from contacts and phylogeny. Specifically, an ancestral chain AB of $2L=200$ spins is evolved along a binary branching tree (see Fig. \ref{fig:phylo_tree}) with $n=10$ generations, yielding $2^{10}=1024$ pairs AB, and $\mu=15$ mutations per branch, under the Hamiltonian in Eq.~\ref{miniHam} on (A) the same Erd\H{o}s-Rényi graph with $p=0.02$ as in Figs.~\ref{fig:partners_mutation} and~\ref{fig:partners_temp}; (B) a stochastic block model graph with two blocks of 100 nodes each, and $p=0.02$ within each block and $p=0.005$ between blocks; (C) an ``interface'' graph with two blocks of 100 nodes each, and $p=0.02$ within each block, but where only 10 nodes in each block are allowed to be in contact with nodes of the other block, with $p=0.25$; (D) a graph corresponding to the contact map from the experimental HK-RR complex structure in Ref.~\cite{Casino09} with threshold at 4 \AA~between closest atoms. These graphs are exactly the same as in Figs.~\ref{fig:magnetisation_against_mutations} and~\ref{fig:histo_magnetisation_graphs}. For each graph, the limiting cases with only contacts and only phylogeny are also shown for comparison. The first one corresponds to independent equilibrium sequences under the Hamiltonian in Eq.~\ref{miniHam} on the same graph. The second one corresponds to neutral evolution on the same binary branching tree. For each graph, each of these three data sets is randomly split into a training set of 400 chains AB and a testing set of 624 chains AB. The latter are randomly divided in 156 species of 4 chains AB each, and pairings between A and B chains are blinded in each species. Partnerships are then predicted using the score in Eq.~\ref{energy}. Each point is averaged over 10 data sets. The null model shows the expectation of the correct prediction fraction if pairs are made randomly within each species. Note that, for sampling, equilibrium is considered reached for 10,000 accepted mutations, except for HK-RR graph where we perform 20,000 accepted mutations (see Fig. \ref{fig:Magnetization}).}
	\label{fig:inference_temp_diff_graphs}
\end{figure}

\begin{figure}[h]
	\centering
	\includegraphics[width=0.9\textwidth]{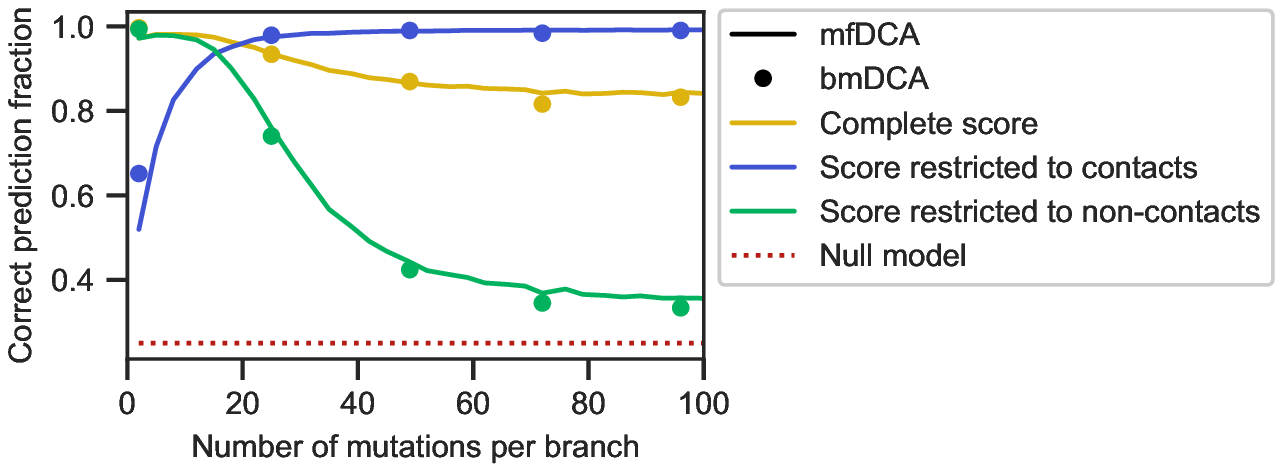}
	\caption{\textbf{Impact of contact and non-contact pairs of sites on partner inference performance using mfDCA or bmDCA.}  The fraction of correctly predicted partner pairs is shown versus the number $\mu$ of mutations per branch of the tree for the minimal model incorporating constraints both from contacts and from phylogeny, either with the full score defined in Eq.~\ref{energy} with couplings inferred using the training set, or with this score restricted to the pairs of sites that are actually in contact (Eq.~\ref{energyb}), or to those that are not in contact (Eq.~\ref{energybb}). Scores are computed using couplings inferred either by mfDCA (as in the rest of this work) or by bmDCA. Note that bmDCA is employed with reweigthing $\theta=0$ here, as is done throughout for mfDCA. Apart from these points, data generation and inference are performed as in Fig.~\ref{fig:partners_mutation}, using the same parameters and the same graph for contacts.  }
	\label{fig:partners_noncontact_bm}
\end{figure}

\begin{figure}[h]
	\centering
	\includegraphics[width=0.9\textwidth]{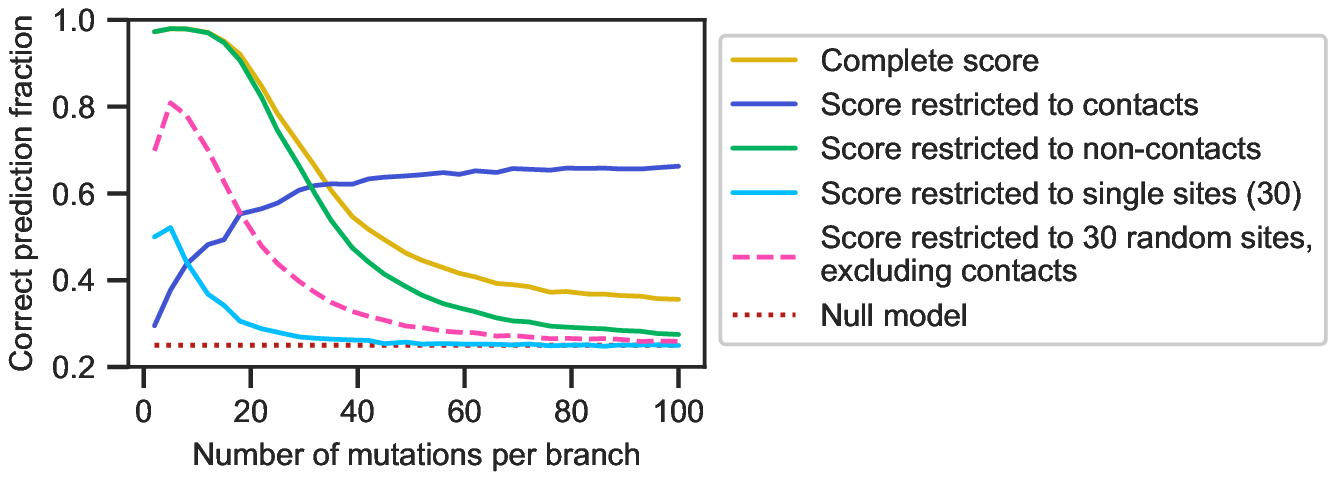}
	\caption{\textbf{Contribution of single (non-connected) sites to partner inference performance.}  The fraction of correctly predicted partner pairs is shown versus the number $\mu$ of mutations per branch of the tree for the minimal model incorporating constraints both from contacts and from phylogeny, either with the full score defined in Eq.~\ref{energy} with couplings inferred using the training set, or with this score restricted to the pairs of sites that are actually in contact (Eq.~\ref{energyb}), or to those that are not in contact (Eq.~\ref{energybb}). In addition, we consider the cases of a score restricted to couplings between those 30 sites in the graph that are not connected to any other site, and of a score restricted to couplings between 30 randomly chosen sites in the graph, excluding couplings between contacting sites. Data generation and inference (apart from the score definition) are performed employing the ``Interface'' graph used in Fig.~\ref{fig:magnetisation_against_mutations} and Fig.~\ref{fig:inference_temp_diff_graphs}C for contacts, at sampling temperature $T=4>T_c$. Apart from these points, data generation and inference are performed as in Fig.~\ref{fig:partners_mutation}. }
	\label{fig:partners_single_interf}
\end{figure}

\begin{figure}[h]
	\centering
	\includegraphics[width=0.9\textwidth]{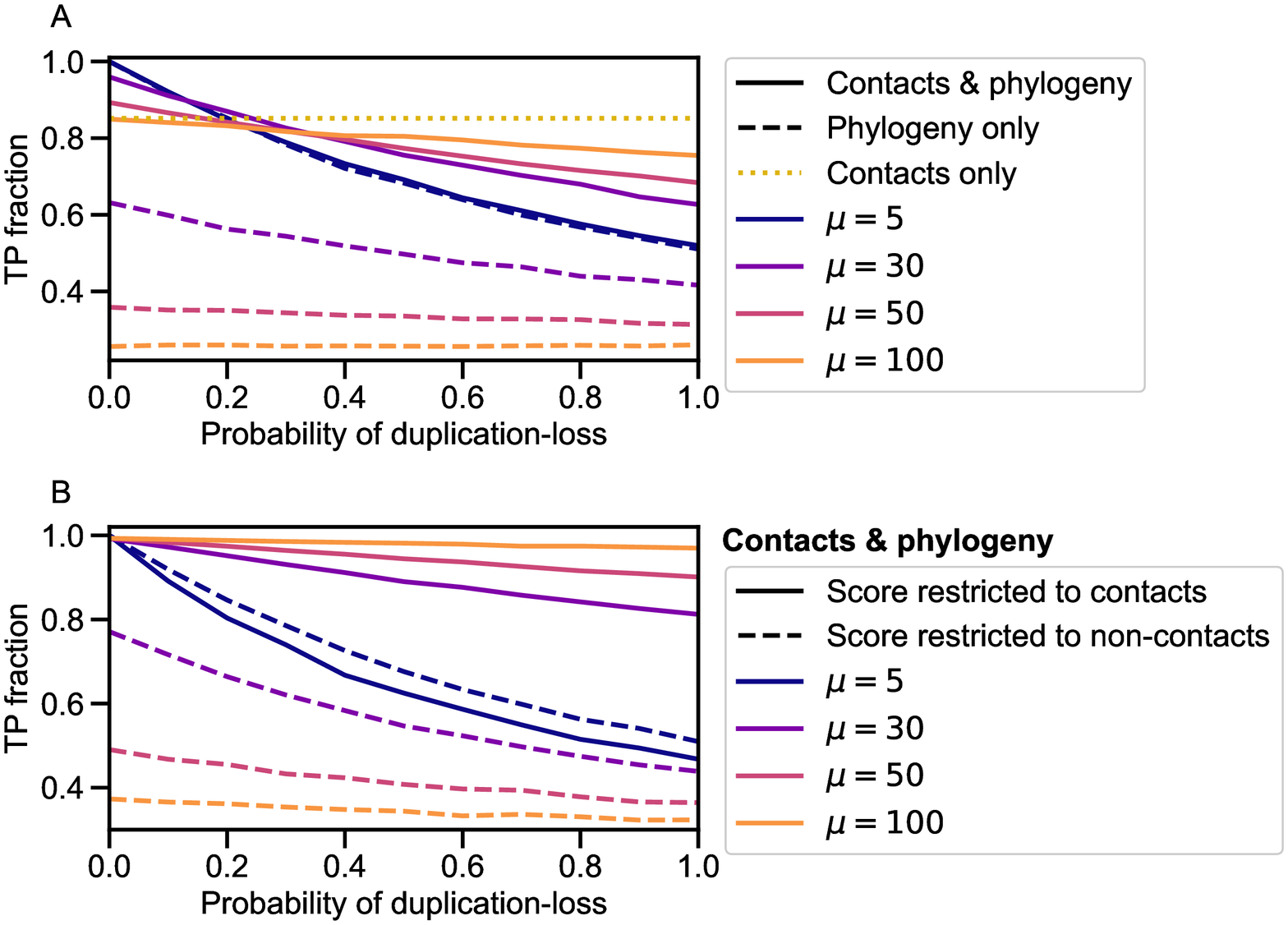}
	\vspace{0.2cm}
	\caption{\textbf{Alternative phylogeny model.}  The fraction of correctly predicted partner pairs is shown versus the fraction of species that undergo a duplication-loss event at each generation in the alternative phylogeny model~\cite{Marmier19}. To generate this data, an ancestral species comprising $m=4$ equilibrium sequence pairs AB sampled at $T=5$ under the Hamiltonian in Eq.~\ref{miniHam} on the same graph as in Fig.~\ref{fig:partners_mutation} was created. Then this ancestral species was evolved by successive speciation events (where all existing species are duplicated) and mutations, performed using the Metropolis criterion with $T=5$ independently for each chain in each species after each speciation event (``generation''). For a certain fraction of speciation events, one sequence AB was randomly removed and one was randomly duplicated in one of the daughter species (duplication-loss event), always keeping $m=4$ (see Ref.~\cite{Marmier19}). Data was generated using a tree of 8 generations, with $\mu$ accepted mutations per sequence and per branch. Because this model starts from one ancestral species with 4 equilibrium chains, this yields 1024 chains AB separated in 256 species. 100 species were randomly selected to form a training set of 400 pairs AB, and the rest constitutes the testing set. Results are averaged over 100 replicates. \textbf{A:} Different values of $\mu$ are considered, both in the model with contacts and phylogeny, and in the one with only phylogeny (where all mutations are accepted). Results are also shown for the case with only contacts, corresponding to a data set of 1024 equilibrium sequences. \textbf{B:} Impact of restricting the pairing scores to the pairs of sites that are actually in contact (Eq.~\ref{energyb}), or to those that are not in contact (Eq.~\ref{energybb}). }
	\label{fig:arbre_alternatif}
\end{figure}

\begin{figure}[h]
	\centering
	\includegraphics[width=0.8\textwidth]{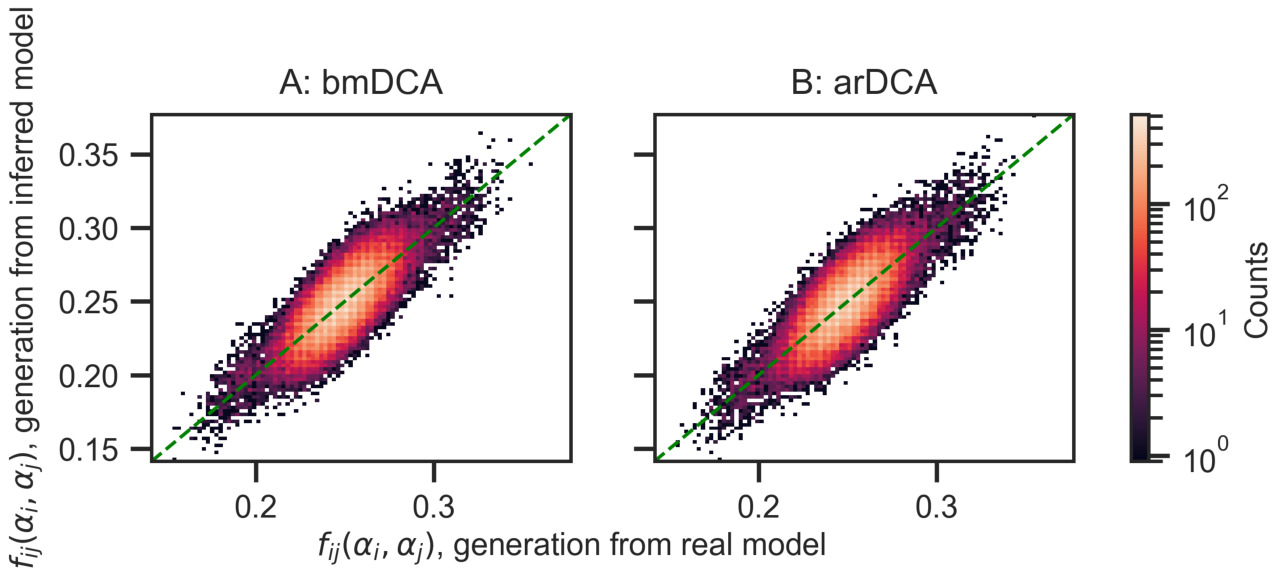}
	\caption{\textbf{Reproduction of two-body frequencies by generative models, for the contact-only minimal model.} An initial data set (``data set 1'') of $1024$ sequences was generated from the contact-only minimal model (Hamiltonian in Eq.~\ref{miniHam} on the same Erd\H{o}s-Rényi graph as in Figs.~\ref{fig:partners_mutation} and~\ref{fig:partners_temp}) at sampling temperature $T=5$. Generative models were inferred from data set 1 by bmDCA (panel A) or arDCA (panel B). A new data set (``data set 2'') of $1024$ sequences was generated, still without phylogeny, from the resulting inferred model. Two-body frequencies in data set 2 are shown versus two-body frequencies in data set 1. The color of each marker represents the number of times it is observed (counts). The green dashed line shows the $y=x$ diagonal. Pearson correlation coefficients are 0.88 and 0.87 for panels A and B, respectively, and linear fits yield intercepts of 0.01 and 0 and slopes of 0.98 and 0.98 for panels A and B, respectively. The two-body frequencies shown for data set 1 include a reweighting of close sequences with Hamming distances under 0.2~\cite{Morcos11, Marks11}, since bmDCA and arDCA aim to match these reweighted frequencies~\cite{Figliuzzi18,Trinquier21}.}
	\label{fig:fij1}
\end{figure}

\begin{figure}[h]
	\centering
	\includegraphics[width=0.8\textwidth]{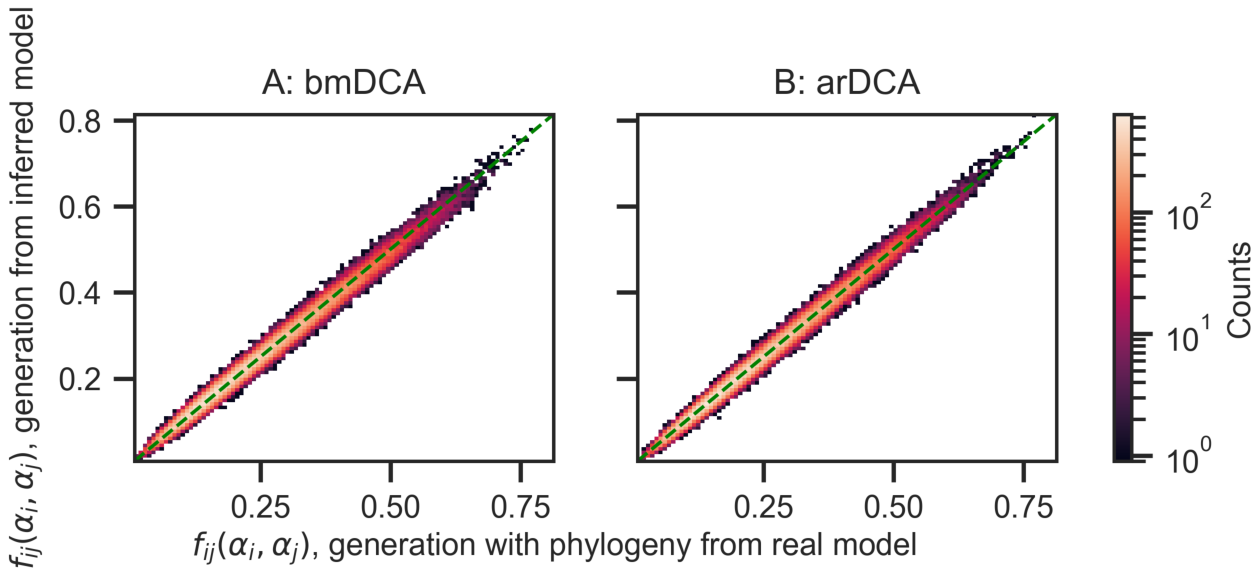}
	\caption{\textbf{Reproduction of two-body frequencies by generative models, for the minimal model with contacts and phylogeny.} An initial data set (``data set 3'') of $1024$ sequences was generated with phylogeny (binary branching tree with $\mu=15$ mutations per branch, and sampling temperature $T=5$) from the minimal model (Hamiltonian in Eq.~\ref{miniHam} on the same Erd\H{o}s-Rényi graph as in Figs.~\ref{fig:partners_mutation} and~\ref{fig:partners_temp}). Generative models were inferred from data set 3 by bmDCA~\cite{Figliuzzi18} (panel A) or arDCA~\cite{Trinquier21} (panel B). A new data set (``data set 4'') of $1024$ sequences was generated without phylogeny from the resulting inferred model. Two-body frequencies in data set 4 are shown versus two-body frequencies in data set 3. The color of each marker represents the number of times it is observed (counts). The green dashed line shows the $y=x$ diagonal. Pearson correlation coefficients are 0.994 and 0.995 for panels A and B, respectively, and linear fits yield intercepts of 0.01 and 0 and slopes of 0.97 and 1 for panels A and B, respectively. The two-body frequencies shown for data set 1 include a reweighting of close sequences with Hamming distances under 0.2~\cite{Morcos11, Marks11}, since bmDCA and arDCA aim to match these reweighted frequencies~\cite{Figliuzzi18,Trinquier21}.}
	\label{fig:fij2}
\end{figure}

\begin{figure}[h]
	\centering
	\includegraphics[width=\textwidth]{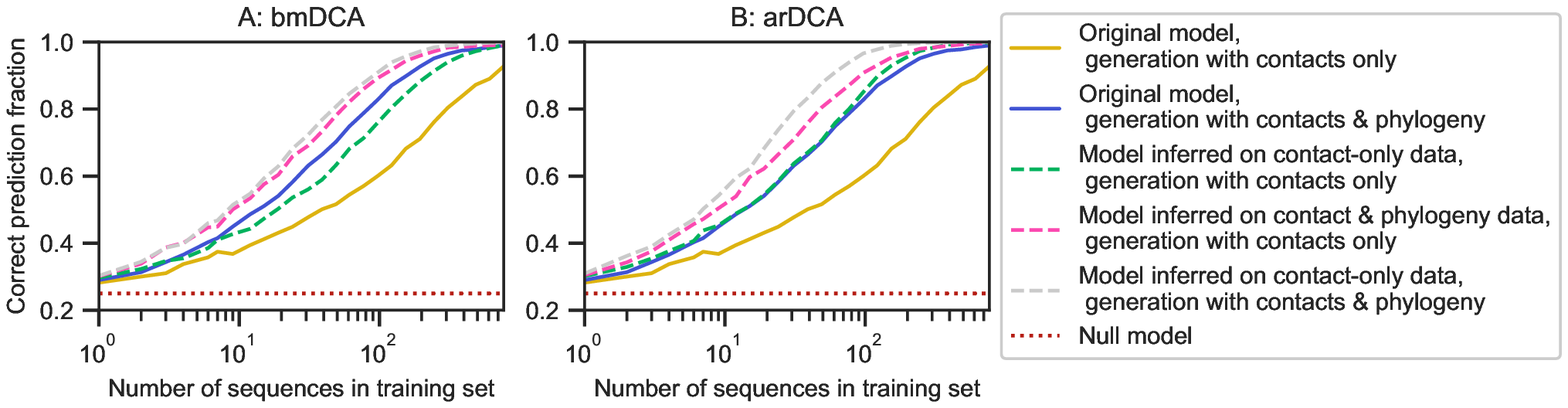}
	\caption{\textbf{Partner inference performance for data generated with original and inferred models.} The fraction of correctly predicted partner pairs is shown versus the number of sequence pairs AB in the training set for data generated under the minimal model, and for data generated using models inferred from this generated data, either by bmDCA~\cite{Figliuzzi18} (panel A) or by arDCA~\cite{Trinquier21} (panel B). Data generation is performed with or without phylogeny. Data generation and partner inference are performed in each case with a variable training set size, exactly as in Fig.~\ref{fig:partners_trainingset}. The data sets generated from the original minimal model correspond to data set 1 described in Fig.~\ref{fig:fij1} (generated with contacts only) and data set 3 described in Fig.~\ref{fig:fij2} (generated with contacts and phylogeny). The data sets generated from inferred models correspond to data sets 2 and 4 described in those figures, and to a data set generated with phylogeny from the model inferred from data set 1. Generation is performed at sampling temperature $T=5$ and with $\mu=15$ mutations per branch in cases with phylogeny. The training set is randomly divided in species of 4 chains AB each. Partnerships are predicted using the score in Eq.~\ref{energy} based on mfDCA. For each point, the inference is performed over 20 generated data sets, and 20 random choices of the training set for each of them. }
	\label{fig:partners_inferc}
\end{figure}

%\clearpage

\begin{figure}[h!]
	\centering
	\includegraphics[width=0.8\textwidth]{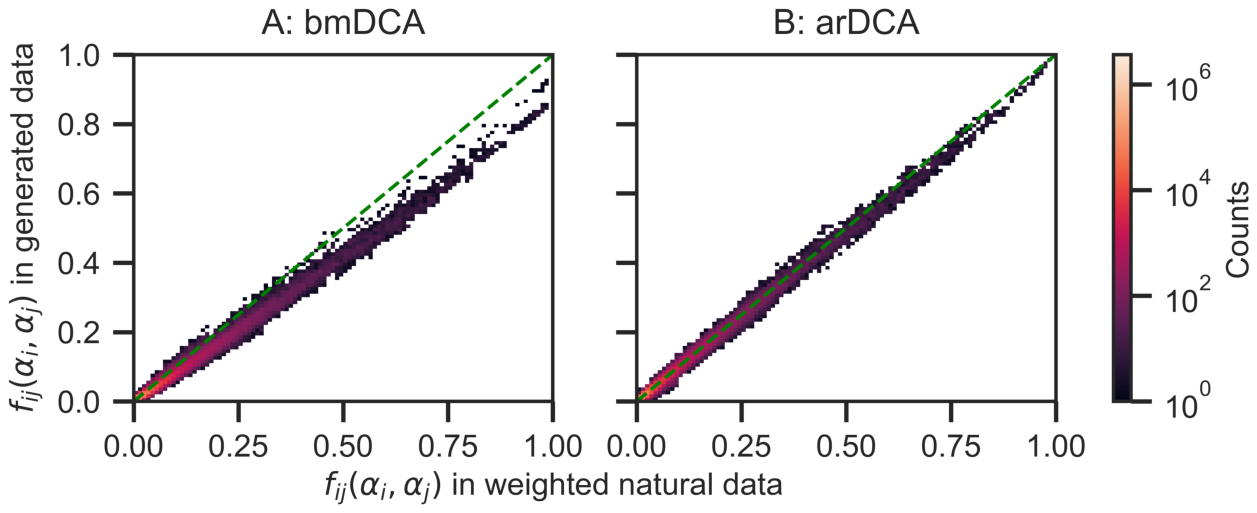}
	\caption{\textbf{Reproduction of two-body frequencies by generative models.} Two body frequencies in a data set of $23,633$ pairs of sequences generated without phylogeny are shown versus the two body frequencies of the natural data set of $23,633$ HK-RR pairs. The color of each marker represents the number of times it is observed (counts). The green dashed line shows the $y=x$ diagonal. Data is generated by bmDCA (panel A) or arDCA (panel B). The associated Pearson correlation coefficients are 0.993 and 0.996 for bmDCA and arDCA, respectively. }
	\label{fig:two-body-freq}
\end{figure}

\begin{figure}[h!]
	\centering
	\includegraphics[width=\textwidth]{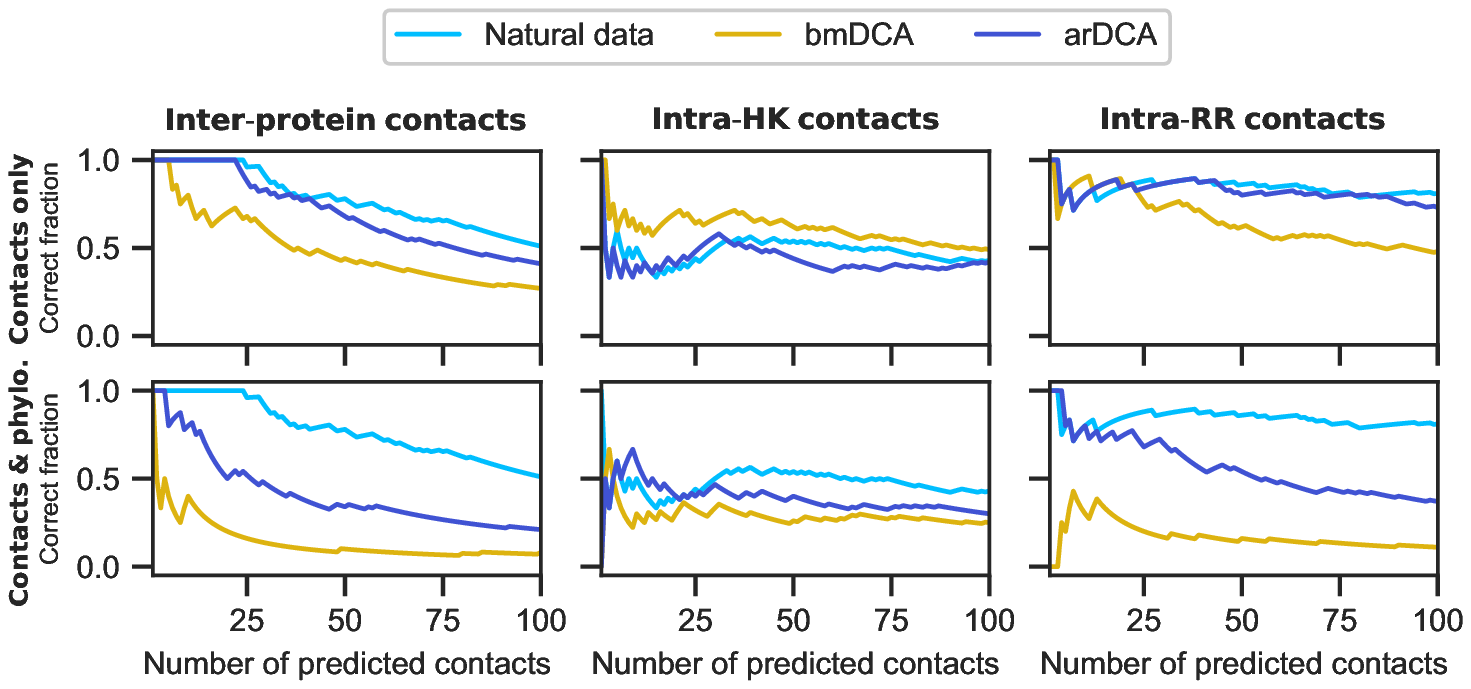}
	\caption{\textbf{Inference of contacts with data generated from models inferred from natural HK-RR data.} The fraction of correctly predicted contacts is shown versus the number of predictions made. A contact between two amino acids is predicted based on the APC-corrected Frobenius norm~\cite{Dunn08,Ekeberg13} of the couplings between amino acids inferred by mfDCA: these scores are ranked from highest to lowest, and the experimental HK-RR complex structure 3DGE~\cite{pdb} is employed to assess whether the contact exists or not in the real protein complex, using a threshold of 8~\AA~between closest atoms. Inter-protein contacts (panels A and D) are distinguished from intra-protein contacts (HK, panels B and E; RR, panels C and F). Inference of contacts is performed on the natural HK-RR data set and on data generated from the model inferred on this natural data set, either by bmDCA or by arDCA. Data is generated either with contacts only or with contacts and phylogeny.}
	\label{fig:21_inference_contact}
\end{figure}

\end{document}